\documentclass[aps,prl,showpacs,english,10pt,a4,nofootinbib,notitlepage,twocolumn,superscriptaddress]{revtex4-1}

\usepackage[utf8]{inputenc}
\usepackage[T1]{fontenc}
\usepackage[english]{babel}  

\usepackage{times}

\usepackage{amssymb,amsmath,amsfonts,amsthm}
\usepackage{mathtools}
\usepackage{verbatim}
\usepackage{enumerate}
\usepackage{graphicx}
\graphicspath{{pics/}}
\usepackage{hyperref}
\usepackage{bbm}
\usepackage{booktabs}
\renewcommand{\arraystretch}{1.5}

\usepackage[caption=false]{subfig}

\usepackage{color}
\definecolor{dred}{rgb}{.8,0.2,.2}
\definecolor{ddred}{rgb}{.8,0.5,.5}
\definecolor{dblue}{rgb}{.2,0.2,.8}
\definecolor{dgreen}{rgb}{.2,0.5,.2}

\theoremstyle{plain}

\theoremstyle{definition}

\newcommand{\bra}[1]{\mbox{$\langle #1|$}}
\newcommand{\ket}[1]{\ensuremath{|#1\rangle}}

\newcommand{\be}{\begin{equation}}
\newcommand{\ee}{\end{equation}}

\def\1#1{{\bf #1}}
\def\2#1{{\cal #1}}
\def\7#1{{\mathbb #1}}

\newcommand{\bea}{\begin{eqnarray}}
\newcommand{\eea}{\end{eqnarray}}

\begin{document}

\title{Simulation of anyonic statistics and its topological path independence using a 7-qubit quantum simulator}

\author{Annie Jihyun Park}
\email{annie.jihyun.park@mpq.mpg.de}
\affiliation{Institute for Quantum Computing and Department of Physics and Astronomy,
University of Waterloo, Waterloo N2L 3G1, Ontario, Canada}
\affiliation{Max-Planck-Institut für Quantenoptik, D-85748 Garching, Germany}
\author{Emma McKay}
\affiliation{Institute for Quantum Computing and Department of Physics and Astronomy,
University of Waterloo, Waterloo N2L 3G1, Ontario, Canada}
\author{Dawei Lu}
\email{d29lu@uwaterloo.ca}
\affiliation{Institute for Quantum Computing and Department of Physics and Astronomy,
University of Waterloo, Waterloo N2L 3G1, Ontario, Canada}
\author{Raymond Laflamme}
\affiliation{Institute for Quantum Computing and Department of Physics and Astronomy,
University of Waterloo, Waterloo N2L 3G1, Ontario, Canada}
\affiliation{Perimeter Institute for Theoretical Physics, Waterloo, Ontario,
Canada}
\affiliation{Canadian Institute for Advanced Research, Toronto, Ontario M5G 1Z8, Canada}

\date{\today}

\begin{abstract}

Anyons, quasiparticles living in two-dimensional spaces with exotic exchange statistics, can serve as the fundamental units for fault-tolerant quantum computation. However, experimentally demonstrating anyonic statistics is a challenge due to the technical limitations of current experimental platforms. Here, we take a state perpetration approach to mimic anyons in the Kitaev lattice model using a 7-qubit nuclear magnetic resonance quantum simulator. Anyons are created by dynamically preparing the ground and excited states of the 7-qubit Kitaev lattice model, and are subsequently braided along two distinct, but topologically equivalent, paths. We observe that the phase acquired by the anyons is independent of the path, and coincides with the ideal theoretical predictions when decoherence and implementation errors are taken into account. As the first demonstration of the topological path independence of anyons, our experiment helps to study and exploit the anyonic properties towards the goal of building a topological quantum computer.

\end{abstract}

\maketitle

\section{\label{intro}Introduction}

It is a fundamental question to investigate the physical properties of exchanging two identical particles. In one-dimensional space, the exchange is trivial as the two particles inevitably collide. In three-dimensional spaces, a wave function of a system acquires either +1 or -1 phase factor for bosons and fermions upon the exchange, respectively. Alternatively, in two-dimensional spaces, exotic statistical properties emerge when exchanging two identical particles, leading to the theoretical existence of anyons. For instance, the wave function can acquire an arbitrary phase factor $e^{i\theta}$, ranging continuously from +1 to -1 for Abelian anyons, or undergo non-trivial unitary evolutions for non-Abelian anyons. This exotic feature of anyons called \emph{fractional statistics} has attracted great interest over the past few decades \cite{khare2005fractional,Tsui1982,moore1991nonabelions,wen1991non,nayak2008non,stern2010non}.

As truly two-dimensional systems do not exist in nature, anyons appear in an effective two-dimensional system as quasi-particles, a collective behaviour of a group of fundamental particles behaving as a single entity.  The experimental evidence of quasi-particle anyons was first discovered in the fractional quantum hall effect in the 1980s \cite{Tsui1982}. Since then, an experimental quest for anyons has interested researchers not only for their fundamentally intriguing feature, but also for their application in performing protected quantum information processing (QIP). As the goal of QIP is to exploit quantum mechanical properties for computation, manipulating quantum properties in a precise manner is critical \cite{ladd2010quantum}. Topological properties of anyons have potential to achieve such a goal, as these properties are resilient to small fluctuations. Therefore, the prospect of utilizing anyons' fractional statistics to achieve robust control has gained much attention. In the past two decades, many powerful quantum computing schemes using anyons have been proposed. Such schemes constitute topological quantum computing (TQC) \cite{sarma2005topologically,nayak2008non,freedman2003topological} and topological quantum error correction \cite{raussendorf2007topological,Kitaev2003}. Of the many proposals, the Kitaev model (KM) of spin lattice \cite{Kitaev2003} is one of the most renowned. By artificially designing a spin lattice model with highly non-trivial ground states, individual localized Abliean anyons can be created and manipulated, leading to the realization of TQC.	

Despite the prospective applications of anyonic statistics, realizing these ideas in experiments remains a challenge, as such tasks typically require generating and manipulating complex many-body quantum systems. Nevertheless, significant progress has been made in small-scale systems theoretically \cite{Han2007,Aguado2008} and experimentally \cite{Pachos2009,Lu2009,du2007experimental,Feng2013}. Through the quantum simulation approach \cite{feynman1982simulating,lloyd1996universal,georgescu2014quantum,friedenauer2008simulating,kim2010quantum,lanyon2010towards,du2010nmr,lanyon2011universal,lu2011simulation,islam2013emergence,li2014experimental}, in which the experimental setup acts as a processor to mimic the dynamics of anyonic systems, several experiments have been implemented to demonstrate the exotic properties of anyons in small systems \cite{Pachos2009,Lu2009,du2007experimental,Feng2013}. These experiments provide better understanding of braiding operations in realistic noise, opening up the possibility of fully utilizing the advantages of anyonic fractional statistics.

However, the path independent nature of the anyons' braiding statistics has not been demonstrated yet as it requires larger quantum simulators with high-fidelity coherent control. In this paper, we study a 7-qubit system with three different paths to braid anyons: two non-trivial paths where the wave function picks up the $\pi$ phase, and one trivial path where the wave function remains unchanged. The additional non-trivial loop allows the experimental proof-of-principle demonstration of path independence (i.e. the statistics do not depend on the shape of the path taken by the anyons as long as the exchange takes place) \cite{Han2007}. This topological feature is one of the key advantages for utilizing anyons. In our experiments, we use a 7-qubit NMR quantum simlulator to realize the three braiding paths through the state preparation approach, and observe that the two phases acquired during the two non-trivial loops agree within experimental uncertainty, although they are below the theoretical value of $\pi$. Our primary source of error  is decoherence that leads to deviations between the theoretically predicted phases and the experimental ones; however, we analyze the errors quantitatively and show that the two non-trivial phases are close to $\pi$ after accounting for such errors.

The remainder of the paper proceeds as follows. In Sec. \ref{Theoretical_Model}, we review the original KM and the simplified 7-qubit KM and describe in detail how anyons gain a $\pi$ phase after braiding regardless of the shapes  of the non-trivial paths. In Sec. \ref{Experimental_Method}, we briefly introduce our experimental setup and the mapping between the KM model and the experimental system, followed by step-by-step description of the implementation in an NMR system. Finally, in Sec. \ref{section:discussion}, we show the experimental results and analyze the errors. We account for experimental deviations from the theoretical predictions using numerical simulations that take realistic error sources into account.
		
\section{Theoretical Model}\label{Theoretical_Model}
\subsection{\label{KM}Kitaev Model}

\begin{figure}[h]%
\centering
\subfloat[Lattice Model]{\includegraphics[scale=0.3]{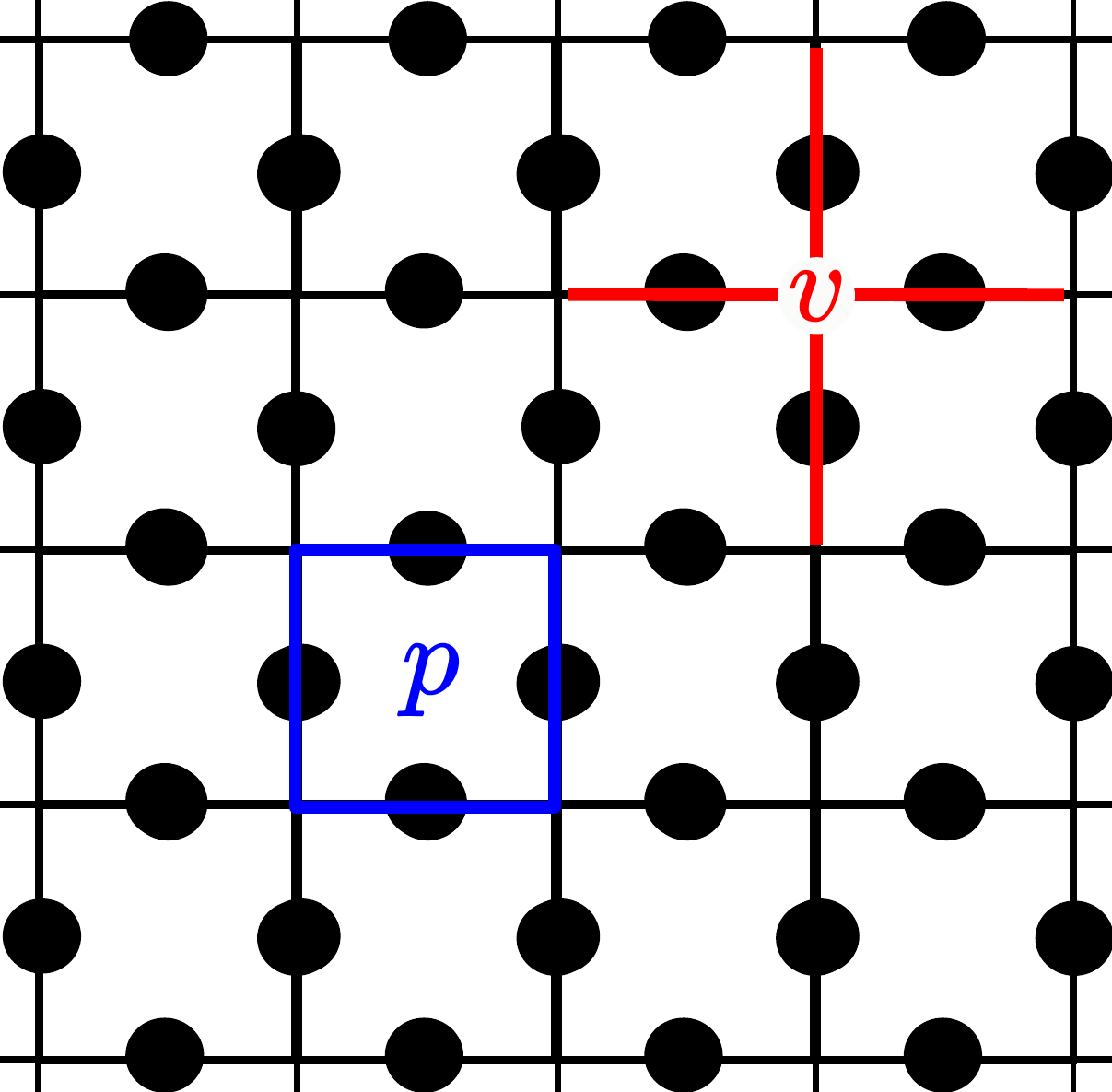}}%
\qquad
\subfloat[Excitations in Lattice]{\includegraphics[scale=0.3]{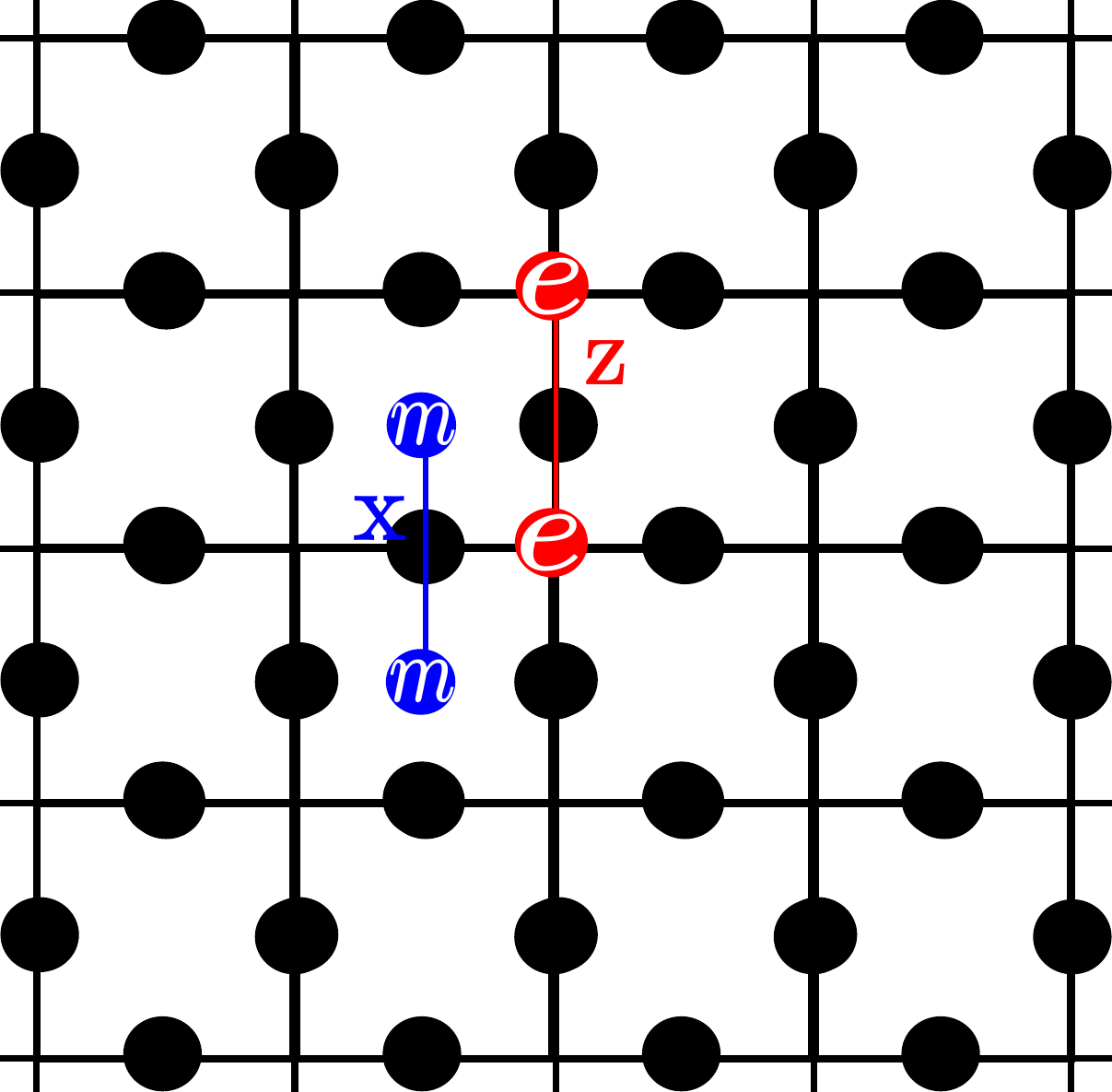}}%
\caption[Kitaev's Lattice Model]{Kitaev model. (a) Qubits are depicted as black dots positioning on the edges of a two-dimensional lattice. There are two different types of four-body interactions $ZZZZ$ and $XXXX$ at a plaquette $p$ ($B_{p}$) and vertex $v$ ($A_{v}$), respectively. (b) Excited states are created by applying single-qubit operators $X$ and/or $Z$. Applying an operator $X_{i}$ changes the eigenvalues of two of the $B_{p}$ operators nearby the $i{\text{th}}$ qubit, thus, shifting the system to a higher energy. Similarly, applying $Z_{i}$ changes the eigenvalues of two of the $A_{v}$ operators nearby the $i{\text{th}}$ qubit. The particular vertices or plaquettes that have been excited by the $i{\text{th}}$ qubit are marked by red and blue dots respectively, and can be identified as having \emph{defects}. These defects can be represented as localized quasi-particle anyons. The anyons on the vertices and plaquettes are called $e$ and $m$ particles, respectively. The excited qubit which creates the pair of `defects' can be identified as a bisector of the string connecting the pair.}%
\label{fig:Kitaev}
\end{figure}
	
In this model, qubits are located on the edges of a two-dimensional lattice as shown in Fig. \ref{fig:Kitaev}(a). The Hamiltonian of the system consists of two different types of four neighbouring-qubit interactions, $XXXX$ and $ZZZZ$ ($X$ and $Z$ are Pauli matrices) at a vertex $v$ and plaquette $p$, respectively. Hence,
\begin{align}
\mathnormal{
H=-\sum_{v}A_{v}-\sum_{p}B_{p},}
\label{eq:Ham_KM}
\end{align}
where $A_{v}=\prod_{j\in \text{star}(v)}X_{j}$ and $B_{p}=\prod_{j\in\text{bond}(p)}Z_{j}$, which are referred to as stabilizer operators. Here, star ($v$) is a set of four spins that share a link with the vertex $v$, and bond ($p$) is a set of four spins placed at the edges of the plaquette $p$. Since all the stabilizer operators commute with each other, the ground state $\ket{\psi_{g}}$ of this Hamiltonian is a +1 eigenstate of the $A_{v}$ and $B_{p}$ operators (note the minus sign in the Hamiltonian in Eq. \ref{eq:Ham_KM}): $A_{v}\ket{\psi_{g}}=\ket{\psi_{g}}$ and $B_{p}\ket{\psi_{g}}=\ket{\psi_{g}}$. The case with a periodic boundary condition on the lattice exemplifies a toric code \cite{Kitaev2003} where the ground states are four-fold degenerate. The degenerate ground states form a protected subspace from possible noise-induced excited states.

\begin{figure}[h]%
\centering
\subfloat[Step 1]{\includegraphics[scale=0.27]{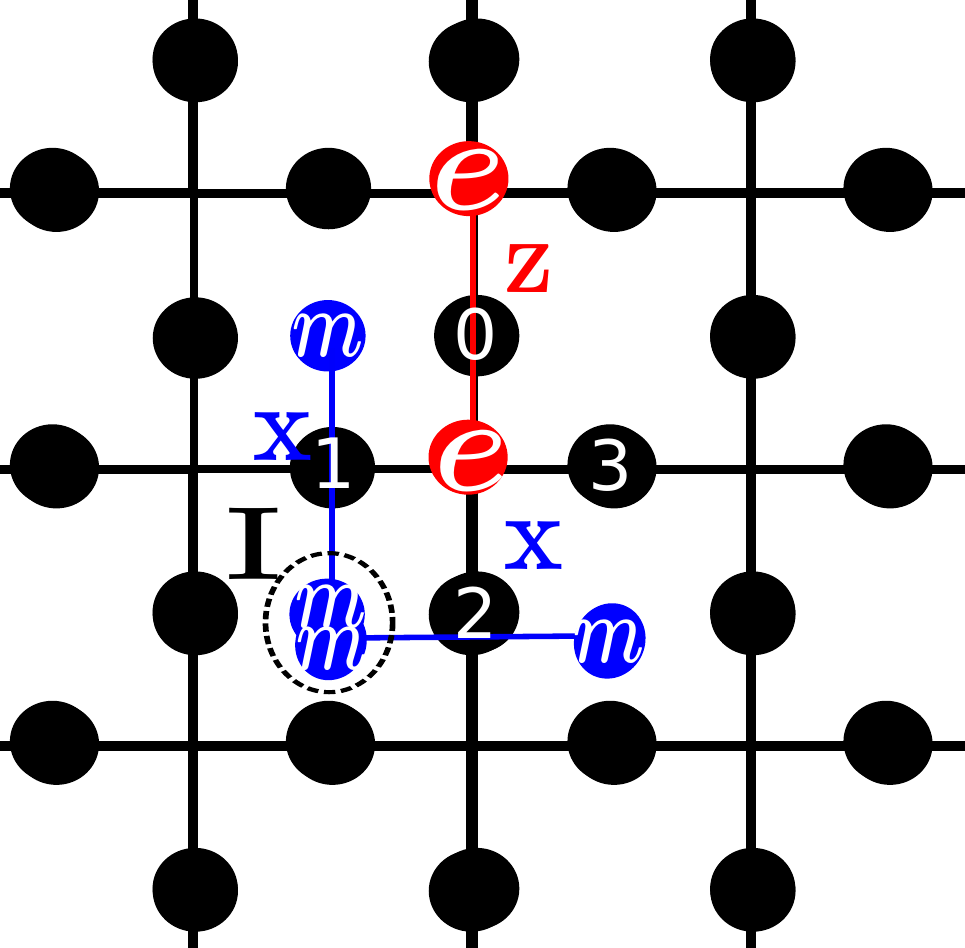}}%
\quad
\subfloat[Step 2]{\includegraphics[scale=0.27]{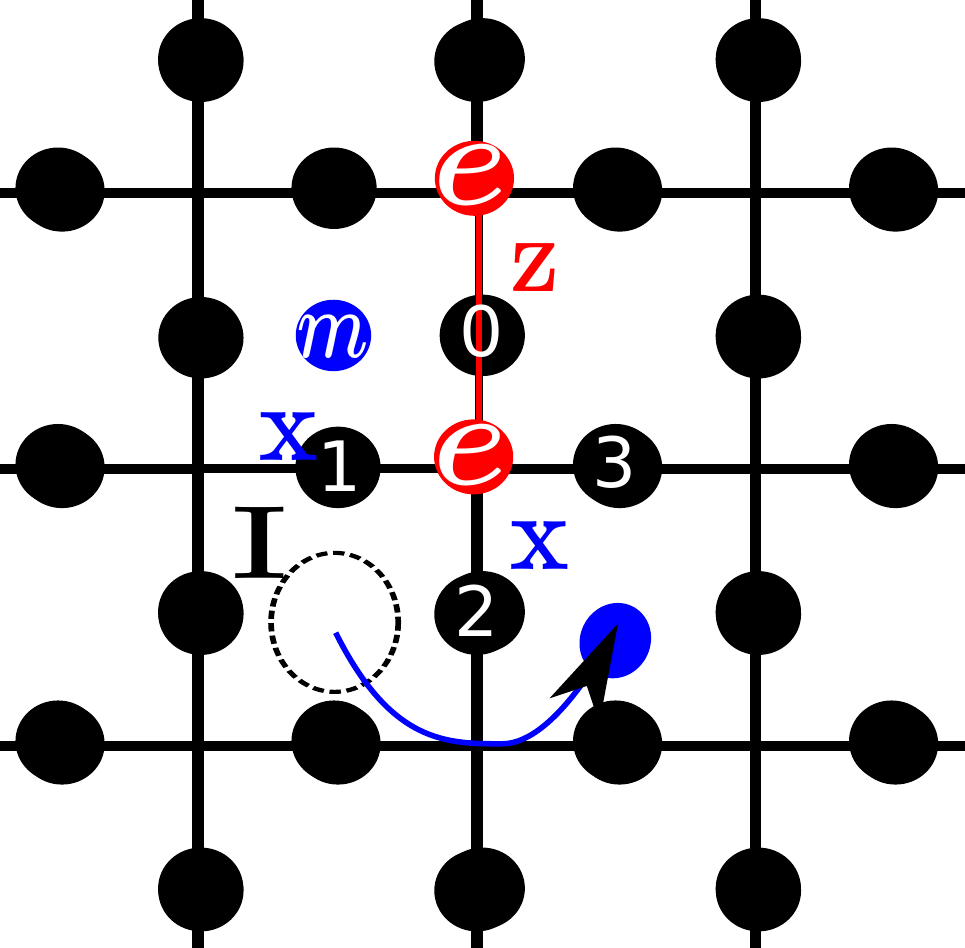}}%
\quad
\subfloat[Step 3]{\includegraphics[scale=0.27]{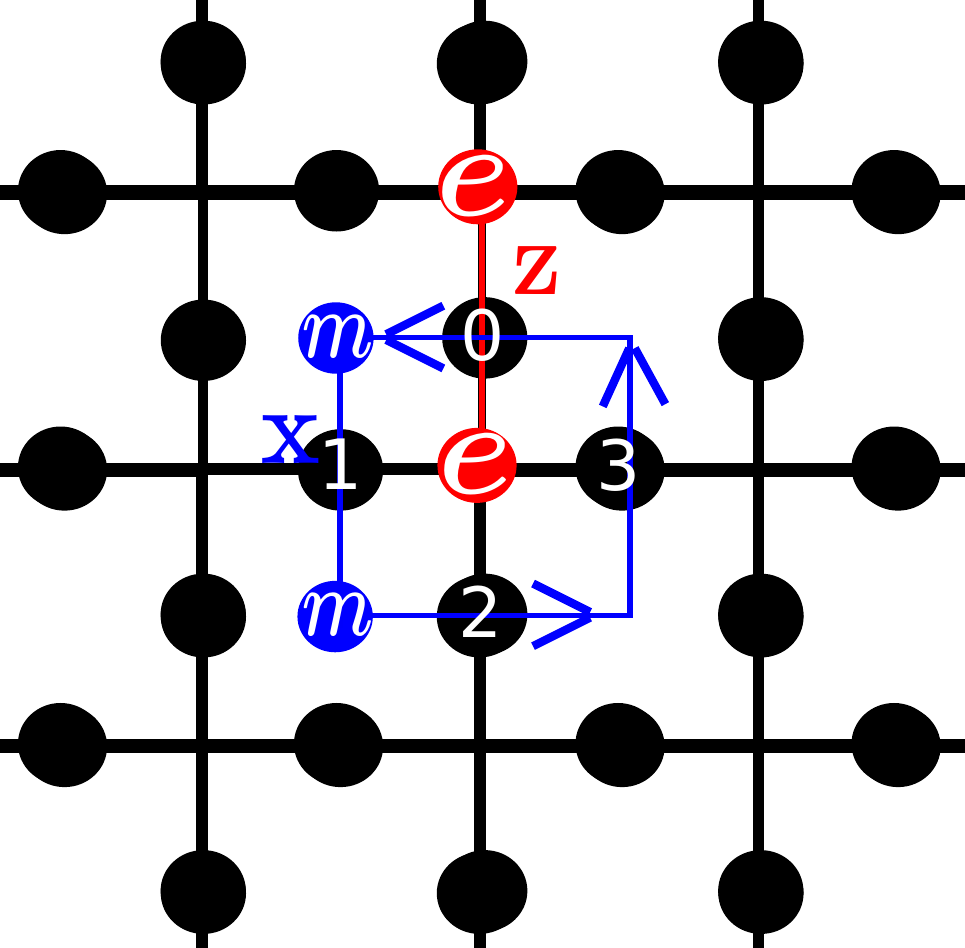}}%
\caption[Kitaev's Lattice Model]{Anyonic braiding operations. A pair of $e$ particles are created by exciting qubit 0 and a pair of $m$ particles are created by exciting qubit 1. (a) Applying $X_{1}$ and $X_{2}$ flips the eigenvalues of their shared $B_{p}$. Hence, creating two defects on the same plquette \emph{annihilates} the defects. (b) Net effect of annihilating the $m$ particles while creating another is moving the particle. (c) Full operation which braids a $m$ particle around $e$ is $X_{0}X_{3}X_{2}X_{1}$. After this braiding operation, the wave function gains a $\pi$ phase. Since such braiding corresponds to exchanging the particles \emph{twice}, this $\pi$ phase demonstrates that the anyonic statistic of $e$ and $m$ particles is $\pi/2$. Note that it is not possible to exchange the two particles' positions once, since they live in different places ($e$ on vertex and $m$ on plquette).}
\label{fig:Kitaev2}
\end{figure}

Excited states of this Hamiltonian are created by applying single-qubit operators $X$ and/or $Z$ to the ground state. These operations create two types of quasiparticles, $e$ particle at a vertex or $m$ particles on a plaquette, as described in Fig. \ref{fig:Kitaev}(b). Subsequently, as shown in Fig. \ref{fig:Kitaev2}(a) and (b), one can move a $m$ particle to a different plaquette by applying $X$ to a qubit nearby. Particles created on the same site annihilate each other so the X operation effectively moves the $m$ particle. Analogously, applying $Z$ to a relevant nearby qubit moves an $e$ particle.
	
One can demonstrate anyonic statistics between $e$ and $m$ particles by moving one around the other, making a closed loop as shown in Fig. \ref{fig:Kitaev2}(c). This braiding operation is equivalent to the two successive particle exchanges. Note that it is not possible to exchange their positions once, since one is located at a vertex and the other at a plaquette. It can be shown that the wave function acquires a -1 phase factor (corresponding to a $\pi$ phase) after such braiding, indicating that a single exchange of Abelian anyon $e$ and $m$ particles would result in a $\pi/2$ phase. Therefore, the anyonic statistics of $e$ and $m$ particles is $\pi/2$.

\subsection{The 7-qubit Kitaev Model}


\begin{figure*}[htb]
\centering
\subfloat{\includegraphics[scale=0.37]{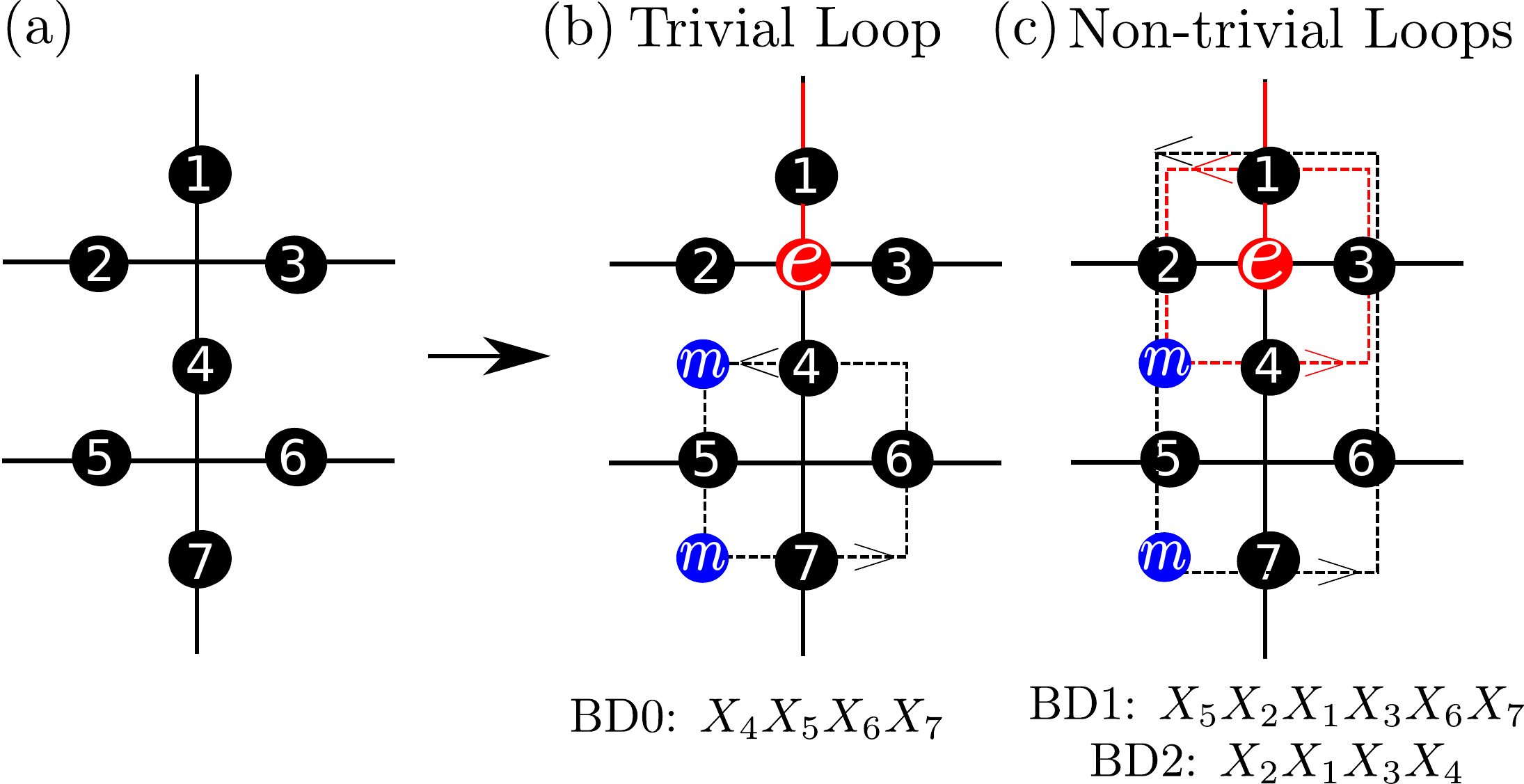}}
\quad\quad\quad
\subfloat{\includegraphics[scale=0.84]{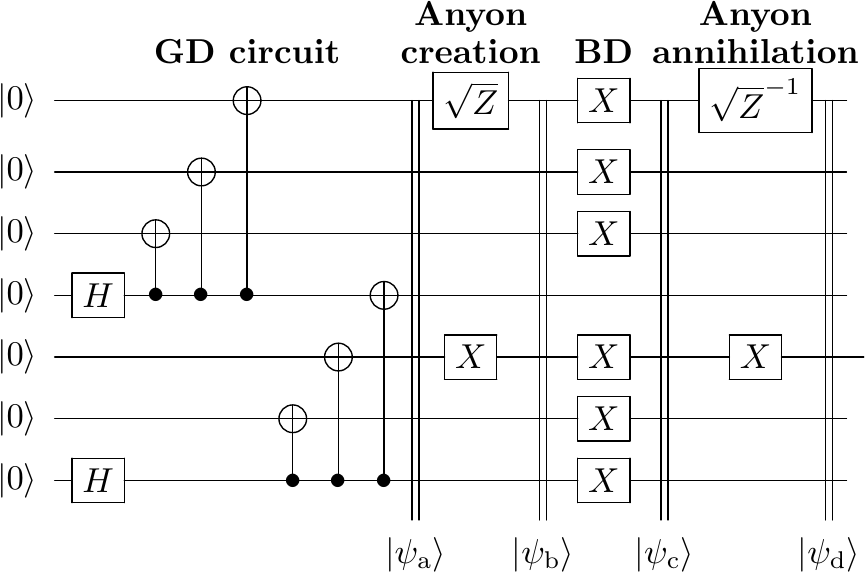}}%

\caption[The 7-qubit Model]{Left: Kitaev model for 7 qubits. (a) Qubits represented by circles are situated on the edges of a 2D lattice, and are subjected to $A_{v}$ and $B_{p}$ interactions shown in Eq. \ref{eq:H}. (b) An $e$ excitation is created by exciting qubit 1, and  two $m$ excitations are created by exciting qubit 5. The black loop, $l_0=X_{4}X_{5}X_{6}X_{7}$, represents a trivial loop in which a $m$ particle is moved along the loop and the wave function of the system remains the same. (c) Black loop $l_1=X_{1}X_{2}X_{3}X_{5}X_{6}X_{7}$ and red loop $l_2=X_{1}X_{2}X_{3}X_{4}$ are non-trivial braiding paths which result in a $\pi$ phase gain of the wave function. Right: Quantum circuit that simulates the anyonic statistics manifested in the 7-qubit KM. The circuit consists of four steps: (i) ground state (GD) circuit which prepares the ground state of the 7-qubit KM $\ket{\psi_{\text{a}}}=\ket{\psi_{g_{7}}}$ from the state $\ket{00\cdots 0}$; (ii) creation of the superposition state $\ket{\psi_{\text{b}}}$ which has two components $\ket{\psi_{mm}}$ and $\ket{\psi_{mme}}$, where $\ket{\psi_{mm}}$ is a state with the pair of $m$ particles, and $\ket{\psi_{mme}}$ is a state with both the $e$ particle and pair of $m$ particles. Such a superposition state is created by applying $\sqrt{Z_{1}}=e^{i\pi/4}(I-iZ_{1})/\sqrt{2}$ and $X_{5}$; (iii) braiding of $e$ and $m$ particles. Without loss of generality, we only show the non-trivial braiding path $l_{1}$ in this circuit. After the braiding, the superposition state $\ket{\psi_{\text{b}}}$ picks up a relative phase on $\ket{\psi_{mme}}$, resulting in the state $\ket{\psi_{\text{c}}}$; (iv) annihilation of anyons, resulting in $\ket{\psi_{\text{d}}}$. When the anyons are braided along the non-trivial paths $l_{1}$ or $l_{2}$, $\ket{\psi_{\text{d}}}$ is $\ket{\psi_{g_{7}}}$, whereas when the path is trivial, $\ket{\psi_{\text{d}}}$ ends up as $\ket{\psi_{e_{7}}}=Z_1\ket{\psi_{g_{7}}}$.}
\label{fig:svmodel}
\end{figure*}	

The 7-qubit model used to demonstrate the path independent property of anyonic braiding is shown in Fig. \ref{fig:svmodel}(a). For the case of a periodic lattice, the Hamiltonian consists of the four body interactions (Eq. \ref{eq:Ham_KM}). However, for the 7-qubit model, we consider a lattice with a rough boudary, which results in two-body $ZZ$ interactions at the boundary. Therefore, the Hamiltonian $H_{7}$ of this system is

\begin{equation}
\label{eq:H}
H_{7}=-A_{1}-A_{2}-B_{1}-B_{2}-B_{3}-B_{4}-B_{5},
\end{equation}
where
\begin{align*}
A_{1}=X_{1}&X_{2}X_{3}X_{4},	\ \
A_{2}=X_{4}X_{5}X_{6}X_{7}, \\
B_{1}=Z_{1}Z_{2},& \ \ B_{2}=Z_{1}Z_{3},\ \
B_{3}=Z_{2}Z_{4}Z_{5},\\
B_{4}=&Z_{3}Z_{4}Z_{6}, \ \ \
B_{5}=Z_{5}Z_{7}.	
\end{align*}

Moreover, due to the absence of periodic boundary conditions, the ground state of this model $\ket{\psi_{g_{7}}}$ is non-degenerate. Since the state $\ket{0}^{\otimes7}$ is already a +1 eigenstate of $B_{1,\cdots ,5}$, the ground state is given by projecting $\ket{0}^{\otimes7}$ on to the +1 eigenstate of $A_{1}$ and $A_{2}$:

\begin{equation}
\label{eq:GD}
\begin{aligned}
\ket{\psi_{g_{7}}}&=\prod_{v=1,2}\frac{1}{\sqrt{2}}(I+A_{v})\ket{0000000} \\
&=\frac{1}{2}(\ket{0000000}+\ket{1111000}+\ket{0001111}+\ket{1110111}).
\end{aligned}
\end{equation}		

Due to the lattice structure, exciting qubit 1 with the operator $Z$ creates a single $e$ particle at $v_{1}$ rather than creating a pair, whereas exciting qubit 5 with the operator $X$ still creates a pair of $m$ particles at the plaquettes associated with $B_{3}$ and $B_{5}$. Refer to Fig. \ref{fig:svmodel}(b) for the particle locations. Starting with this excited state, there are three possible loops to braid $m$ particles as shown in Fig. \ref{fig:svmodel}(b,c): the trivial braiding operation $l_0=X_{4}X_{5}X_{6}X_{7}$ where a $m$ particle braids around $v_{2}$ without an $e$ particle, and the two non-trivial braiding operations $l_1=X_{1}X_{2}X_{3}X_{5}X_{6}X_{7}$ and $l_2=X_{1}X_{2}X_{3}X_{4}$ where a $m$ particle braids around $v_{1}$ with an $e$ particle. The wave function remains the same when the braiding operation is trivial; however, if the operation is non-trivial, the wave function picks up a $\pi$ phase from the fractional statistics.
	
To experimentally demonstrate the path independence of anyonic braiding, we simulated anyonic physics manifested in the 7-qubit KM using a liquid-state NMR quantum simulator. Since it is experimentally challenging to engineer the KM Hamiltonian which involves four-body interactions, we took the state preparation approach: dynamically preparing the ground and excited states of the KM Hamiltonian in a NMR system, instead of generating the KM Hamiltonian and cooling the system.

The quantum circuit which simulates the anyonic physics is shown on the right in Fig. \ref{fig:svmodel}. The main idea is to prepare $\ket{\psi_{g_{7}}}$ and then create a superposition $(\ket{\psi_{mm}}+\ket{\psi_{mme}})/\sqrt{2}$, where $\ket{\psi_{mm}}$ is a state with the pair of $m$ particles, and $\ket{\psi_{mme}}$ is a state with both the $e$ particle and pair of $m$ particles. If braided along the non-trivial paths such as $l_1$ or $l_2$ in which $e$ circulates around $m$, $\ket{\psi_{mme}}$ gains a $\pi$ phase due to the fractional statistics; otherwise, $\ket{\psi_{mme}}$ remains unchanged. By measuring the variation of the relative phase on $\ket{\psi_{mme}}$ before and after the braiding, one can deduce whether the braiding path is trivial or not and ,furthermore, demonstrate the path independence. The details are described as follows.

First, two Hadamard and six controlled-NOT (CNOT) gates are applied to prepare the ground state $\ket{\psi_{g_{7}}}$ of the 7-qubit KM Hamiltonian from $\ket{00\cdots 0}$, as depicted in Fig. \ref{fig:svmodel}. Then, applying $X_5$ and $Z_{1}X_{5}$ on  $\ket{\psi_{g_{7}}}$ generates
\be
\ket{\psi_{mm}} = X_5\ket{\psi_{g_{7}}}, \ket{\psi_{mme}} = Z_{1}X_{5}\ket{\psi_{g_{7}}},
\ee
respectively. To create a superposition of the two, we apply $\sqrt{Z_{1}}X_{5}$ since $\sqrt{Z_{1}}=e^{i\pi/4}(I-iZ_{1})/\sqrt{2}$. When the anyons are braided along a non-trivial loop, the superposition picks up a relative phase on the $\ket{\psi_{mme}}$ component. Finally, anyons are annihilated by reversing the creation operator $\sqrt{Z_{1}}X_{5}$ in order to measure this relative phase. The system ultimately evolves to either the ground state  $\ket{\psi_{g_{7}}}$ or the excited state $\ket{\psi_{e_{7}}}$ depending on different braiding paths. Therefore, we can experimentally demonstrate the path independence nature of anyonic braiding if the two phases obtained under the two different non-trivial loops $l_{1}$ and $l_{2}$ are the same.

The states corresponding to each step of the circuit shown on the right in Fig. \ref{fig:svmodel} are
\begin{align}
\ket{\psi_{\text{a}}}&=\ket{\psi_{g_{7}}}, \\
\ket{\psi_{\text{b}}}&=\sqrt{Z_{1}}X_5 \ket{\psi_{\text{a}}} \nonumber \\
&=\frac{e^{i\pi/4}}{\sqrt{2}}(\ket{\psi_{mm}}-i\ket{\psi_{mme}}), \\
\ket{\psi_{\text{c}}}&=l_{0,1,2} \ket{\psi_{\text{b}}} \nonumber \\
&=\frac{e^{i\pi/4}}{\sqrt{2}}(\ket{\psi_{mm}}-ie^{i\theta_a}\ket{\psi_{mme}}), \\
\ket{\psi_{\text{d}}}&=\sqrt{Z_{1}}^{-1}X_5 \ket{\psi_\text{c}} \nonumber \\
&=\frac{1}{2}((1+e^{i\theta_a})\ket{\psi_{g_7}}+i(1-e^{i\theta_a})Z_1\ket{\psi_{g_7}}) \nonumber \\
&=\frac{1}{2}((1+e^{i\theta_a})\ket{\psi_{g_7}}+i(1-e^{i\theta_a})\ket{\psi_{e_7}}), \label{eq:psi d}
\end{align}
where $\theta_{a}$ is the phase gained from the anyonic statistics for different loops $l_{0,1,2}$. When the $m$ moves around the trivial loop $l_{0}$, the final state $\ket{\psi_{\text{d}}}$ ends up at the ground state $\ket{\psi_{g_{7}}}$ ($\theta_{a}=0$), whereas when the $m$ is moved around the non-trivial loops $l_{1}$ or $l_{2}$, $\ket{\psi_{\text{d}}}$ ends up at the excited state $\ket{\psi_{e_{7}}}$ ($\theta_{a}=\pi$). In order to demonstrate path independence of anyonic braiding experimentally, we need to implement the entire circuit and observe $\theta_{a}$ for different loops.

\section{Experimental Implementation in NMR}
\label{Experimental_Method}
\begin{figure*}[htb]%
\begin{center}
\includegraphics[width= 2\columnwidth]{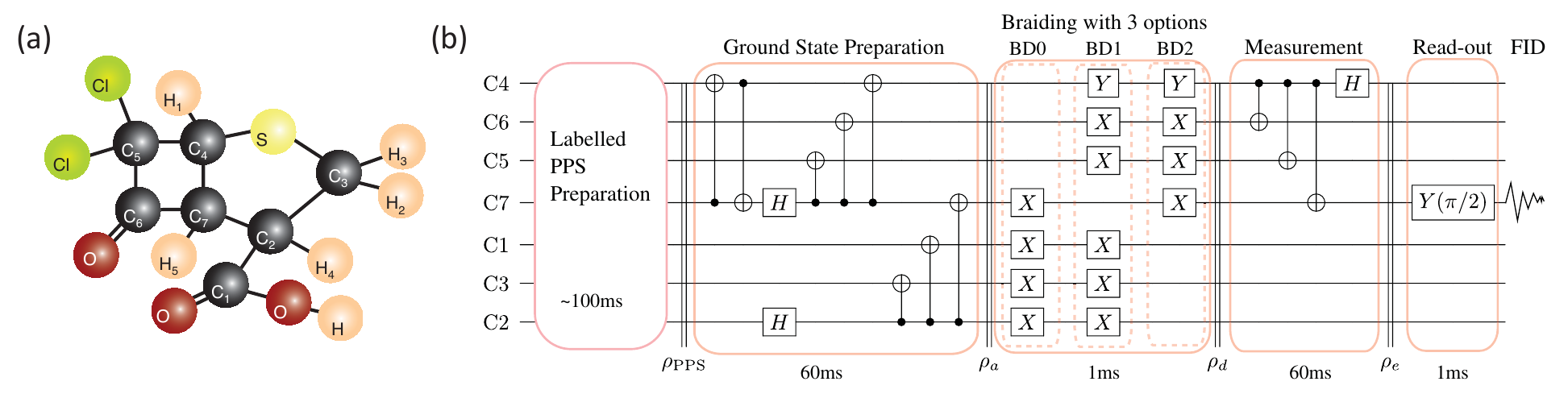}%
\end{center}
\setlength{\abovecaptionskip}{-0.3cm}
\caption{(a) Molecular structure of per-$^{13}$C-labeled dichlorocyclobutanone derivative, where C$_1$ to C$_7$ form a 7-qubit system. (b) Schematic NMR circuit showing the overview of the experimental scheme. It contains five steps with the sequence length at the bottom of each step: the labeled PPS state preparation based on the cat-state method \cite{Knill2000}, and the detailed network can be found in the Appendix A; preparation of the ground state $\ket{\psi_{g_{7}}}$ of the KM Hamiltonian; anyonic manipulation including creation, braiding, and annihilation (all three braiding paths are shown here, but in each experiment we just implement one); measurement circuit which converts the state tomography to simpler diagonal elements measurement; readout pulse on C$_{7}$ to measure the required diagonal elements.}
\label{fig:NMR sample}
\end{figure*}

Our 7-qubit NMR processor is the per-$^{13}$C-labeled dichlorocyclobutanone  derivative \cite{johnson2008cyclobutanone,lu2015experimental} dissolved in
d6-acetone. The molecule consists of seven $^{13}$C spins and the five $^{1}$H spins. We denoted the seven nuclear spins of $^{13}$C as qubits, while $^1$H nuclei were decoupled throughout all experiments except for the initialization step to boost polarization on $^{13}$C. The molecular structure is depicted in Fig. \ref{fig:NMR sample}(a), where two nearest-neighbouring $^{13}$Cs have stronger coupling strengths, implying the ability to implement a faster two-qubit gate. Therefore, by comparing the geometry of KM qubits and the structure of nuclear spins, we mapped each KM qubit to the nuclear spin in as shown Fig. \ref{fig:NMR sample}(b). The natural Hamiltonian of this system is described as
\begin{align}
\mathcal{H}_{\text{NMR}}=\sum_{i}^{7}\frac{\nu_{i}}{2}Z_{i}+\sum_{i<j}\frac{J_{ij}}{4}Z_{i}Z_{j},
\label{eq:Hnmr}
\end{align}
where $\nu_{i}$ is the chemical shift frequency of the $i{\text{th}}$ spin, and $J_{ij}$ is the coupling strength between the  $i{\text{th}}$ and $j{\text{th}}$ spins (refer to Appendix A for values of the parameters). All experiments were conducted on a Bruker DRX 700 MHz spectrometer at room temperature. The experiment was divided into five main steps as shown in Fig. \ref{fig:NMR sample} (b), as follows:

\emph{PPS initialization.} We first utilized the cat-state method proposed in \cite{Knill2000} to initialize the system to a labeled pseudo-pure state (PPS) state. It can be represented by a deviation matrix of the form $\widetilde{\rho}_{\text{PPS}}=\ket{000000}\bra{000000}Z_{\text{C}_{7}}$,
where C$_7$ is the labeling qubit. Two techniques were adopted before this initialization step to improve the signal-to-noise (SNR) ratio. One is turning on $^{13}$C and $^1$H couplings temporarily at the very beginning and applying a SWAP gate between C$_7$ and H$_5$, to achieve a $\sim$4 times higher polarization on C$_7$. The other one is performing \emph{RF-selection} \cite{Ryan2009} sequence to pick out a slice of the NMR sample which experiences much better radio-frequency (RF) homogeneity by randomizing the other part with worse RF homogeneity. Subsequently, the labeled PPS state was prepared using non-unitary transformations via gradient fields and phase cycles \cite{Knill2000}. The total length of the initialization sequence is about 100 ms. Refer to Appendix A for more details about the
PPS initialization step.

\emph{Ground state preparation.} Unlike the theoretical circuit on the right in Fig. \ref{fig:svmodel}, the implemented circuit in NMR prepared the ground state from $\widetilde{\rho}_{\text{PPS}}=\ket{000000}\bra{000000}Z_{\text{C}_{7}}$, rather than the required pure state $\ket{0000000}$. Nevertheless, since $\widetilde{\rho}_{\text{PPS}}$ contains half of  $\ket{0000000_{\text{C}_{7}}}$ and half of $\ket{0000001_{\text{C}_{7}}}$, we can simply write the deviation matrix after the ground state preparation as:
\begin{align}
\widetilde{\rho}_{g_{7}}=\frac{1}{2}\left(\ket{\psi_{g_{7}}}\bra{\psi_{g_{7}}}-\ket{\psi_{\widetilde{g_{7}}}}\bra{\psi_{\widetilde{g_{7}}}}\right)
\end{align}
where $\ket{\psi_{\widetilde{g_{7}}}}$ results from $\ket{0000001_{\text{C}_{7}}}$. Under perfect unitary transformation, $\ket{\psi_{\widetilde{g_{7}}}}\bra{\psi_{\widetilde{g_{7}}}}$ stays orthogonal to $\ket{\psi_{g_{7}}}\bra{\psi_{g_{7}}}$ throughout the circuit, thus not interfering with the final result if it can be separated in the NMR spectra. In fact, the additional two CNOT gates in the beginning of the ground state preparation shown in Fig. \ref{fig:NMR sample}(b) were specifically added to achieve this separation. However, in the presence of errors, $\ket{\psi_{\widetilde{g_{7}}}}\bra{\psi_{\widetilde{g_{7}}}}$ did slightly modify with the final result, as analyzed in Sec. \ref{section:discussion}.

 The entire ground state preparation step was optimized by a 60 ms GRadient Ascent Pulse Engineering (GRAPE) pulse \cite{Khaneja2005} based on a subspace approach \cite{ryan2008}. The simulated fidelity of this pulse is over 0.99. Additionally, a special rectification method was used in the experiment to ensure that all of the GRAPE pulses acting on the spins were very close to theoretical expectations \cite{weinstein2004,moussa2012}. We performed modified  stabilizer measurements after the ground state preparation step to verify the state. This step is explained in detail in Appendix \ref{appen:gd}.

\emph{Anyon creation, braiding and annihilation.} These three parts shown in the emulation circuit (on the right of Fig. \ref{fig:svmodel}) are compressed together to simplify the circuit as they only involve single-qubit rotations. The trivial loop $l_{0}$ and non-trivial loops $l_{1}$, $l_{2}$ are all depicted in Fig. \ref{fig:NMR sample}(b), and in each experiment only one loop was implemented. The three braiding operators were realized by 1 ms GRAPE pulses, respectively. In principle, after this stage we can determine $\theta_a$ by measuring coefficients of $\ket{\psi_{g_{7}}}$ and $\ket{\psi_{e_{7}}}$ in Eq. \ref{eq:psi d}, but it does require many measurements in a 7-qubit system.

\emph{Measurement.} This additional `measurement' step is added to estimate $\theta_{a}$ with a few measurements, which allows us to measure diagonal elements of the final state and then extract the value of $\theta_a$. It separates diagonal elements of $\ket{\psi_{g_{7}}}$ and $\ket{\psi_{e_{7}}}$ via basis transformation by evolving the state $\ket{\psi_{\text{d}}}$ to
\begin{align}
\ket{\psi_{\text{e}}}&=\frac{1}{2\sqrt{2}}\bigg((1+e^{i\theta})(\mathrm{\ket{0000000}} +\mathrm{\ket{0001111}})\\ \nonumber
&+i(1-e^{i\theta})(\mathrm{\ket{1000000}} +\mathrm{\ket{1001111}})\bigg). 			
\end{align}
Therefore, considering $\ket{\psi_{\widetilde{g_{7}}}}\bra{\psi_{\widetilde{g_{7}}}}$, the final density matrix is
\begin{align}
\widetilde{\rho}_{\text{e}}&=\frac{1}{2}\left(\ket{\psi_{\text{e}}}\bra{\psi_{\text{e}}}-\ket{\psi_{\widetilde{\text{e}}}}\bra{\psi_{\widetilde{\text{e}}}}\right) \\ \nonumber &=\frac{1}{2}(|\alpha|^{2}\ket{\psi_{\text{p0}}}\bra{\psi_{\text{p0}}}+\alpha\beta^{*}\ket{\psi_{\text{p0}}}\bra{\psi_{\text{p1}}}+\alpha^{*}\beta\ket{\psi_{\text{p1}}}\bra{\psi_{\text{p0}}} \\ \nonumber
&+|\beta|^{2}\ket{\psi_{\text{p1}}}\bra{\psi_{\text{p1}}}-\ket{\psi_{\widetilde{\text{e}}}}\bra{\psi_{\widetilde{\text{e}}}}),
\end{align}
with
\begin{align}
\ket{\psi_{\text{p0}}}& =  \ket{0000000}+\ket{0001111}, \\
\ket{\psi_{\text{p1}}}& = \ket{1000000}+\ket{1001111}.
\end{align}
The coefficients $\alpha=(1+e^{i\theta})/2\sqrt{2}$, and $\beta=i(1-e^{i\theta})/2\sqrt{2}$, and $\ket{\psi_{\widetilde{\text{e}}}}\bra{\psi_{\widetilde{\text{e}}}}$ originates from the neglected part $\ket{\psi_{\widetilde{g_{7}}}}\bra{\psi_{\widetilde{g_{7}}}}$. In this case,
\begin{align}
\theta_{a}=2\arctan(\sqrt{|{\frac{\beta}{\alpha}}|^{2}}), \ \ \ \ \ -\pi < \theta < \pi.
\label{eq:phase}
\end{align}
To evaluate $\theta_{a}$, we estimated $|\alpha|^{2}$ by measuring the diagonal elements of $\ket{\psi_{\text{p0}}}\bra{\psi_{\text{p0}}}$ and similarly $|\beta|^{2}$ by measuring the diagonal elements of $\ket{\psi_{\text{p1}}}\bra{\psi_{\text{p1}}}$.

\emph{Diagonal elements readout.} Since the diagonal elements cannot be directly observed in NMR, we indirectly measured them by applying the readout pulse which rotates C$_7$ by $\pi$/2 around the $y$-axis. This readout pulse generated single coherences from the diagonal elements, and thus a detectable signal with distinct frequencies depending on the state of the other qubits (see Appendix for detailed descriptions). In particular, the transitions relevant to $|\alpha|^{2}$ and $|\beta|^{2}$ estimations are at four distinct frequencies centered around $\nu_{7}$ (resonant frequency of C$_7$): 61.25Hz, 24.09Hz, 32.24Hz, and -4.93Hz. Therefore, the real coefficients of the peaks at these specified frequencies can indirectly estimate the diagonal elements of interest.

There is one assumption in the measurement of diagonal elements in the above method. The peaks are actually generated by the subtraction of two relevant diagonal elements after rotating C$_7$ by $\pi$/2 around the $y$-axis (see Appendix). For example, the intensity of the peak at 61.25Hz corresponds to $\ket{0000000}\bra{0000000}-\ket{0001000}\bra{0001000}$, but we only need the value of the first term. So we assume that the latter term is 0 in order to get the value of the first term. We simulated the contributions from such small elements and found that this assumption should be good enough for the accurate estimation of $\theta_{a}$.

\begin{table*}[htb]
\setlength{\tabcolsep}{5pt}
\renewcommand{\arraystretch}{1.3}
\small
\vspace*{-0.28cm}
    \begin{tabular}{|c|c|c|c|c|c|c|} \cline{1-7}
       & \multicolumn{2}{c|}{${|\alpha|^{2}}$}  & \multicolumn{2}{c}{${|\beta|^{2}}$} & \multicolumn{2}{|c|}{${\theta_{a}}$}  \\ \cline{2-7}
			& theory & experiment & theory & experiment & theory & experiment \\  \cline{1-7}
      \textbf{No BD} & 1 &   0.83$\pm$0.01 &0& 0.01$\pm$0.01 & 0 & (12.1$\pm$9.5)$^{\circ}$ \\ \cline{1-7}
      \textbf{BD0}   & 1 &   0.83$\pm$0.01 &0& 0.02$\pm$0.01 & 0 & (17.4$\pm$6.0)$^{\circ}$\\  \cline{1-7}
      \textbf{BD1}   & 0 &   0.05$\pm$0.01 &1& 0.85$\pm$0.01 &$\pi$ (180$^{\circ}$) & (153.9$\pm$3.8)$^{\circ}$\\  \cline{1-7}
      \textbf{BD2}   & 0 & 0.05$\pm$0.01& 	1& 0.81$\pm$0.02 & $\pi$ (180$^{\circ}$) & (151.4$\pm$3.8)$^{\circ}$  \\ \cline{1-7}
    \end{tabular}
\vspace{0.2cm}
\caption{Experimentally evaluated $|\alpha|^{2}$, $|\beta|^{2}$ and $\theta_{a}$ values compared with the theoretical values. $|\alpha|^{2}$ and $|\beta|^{2}$ are evaluated by fitting the intensities of the peaks at the frequencies of $a$ and $b$, and frequencies $c$ and $d$, respectively. Subsequently, the anyonic phases $\theta_{a}$ are determined via Eq. \ref{eq:phase}. $|\alpha|^{2}$ and $|\beta|^{2}$ are normalized to be in the range of $0\leq|\alpha|^{2}$, $|\beta|^{2}\leq 1$. The standard deviations of $|\alpha|^{2}$, $|\beta|^{2}$ are the fitting errors, and the standard deviations of $\theta_{a}$ are calculated from $|\alpha|^{2}$, $|\beta|^{2}$ using the error propagation method.}
\label{table:phase}
\end{table*}

\section{Result and Discussion}\label{section:discussion}

\begin{figure*}[htb]
\includegraphics[scale=0.45]{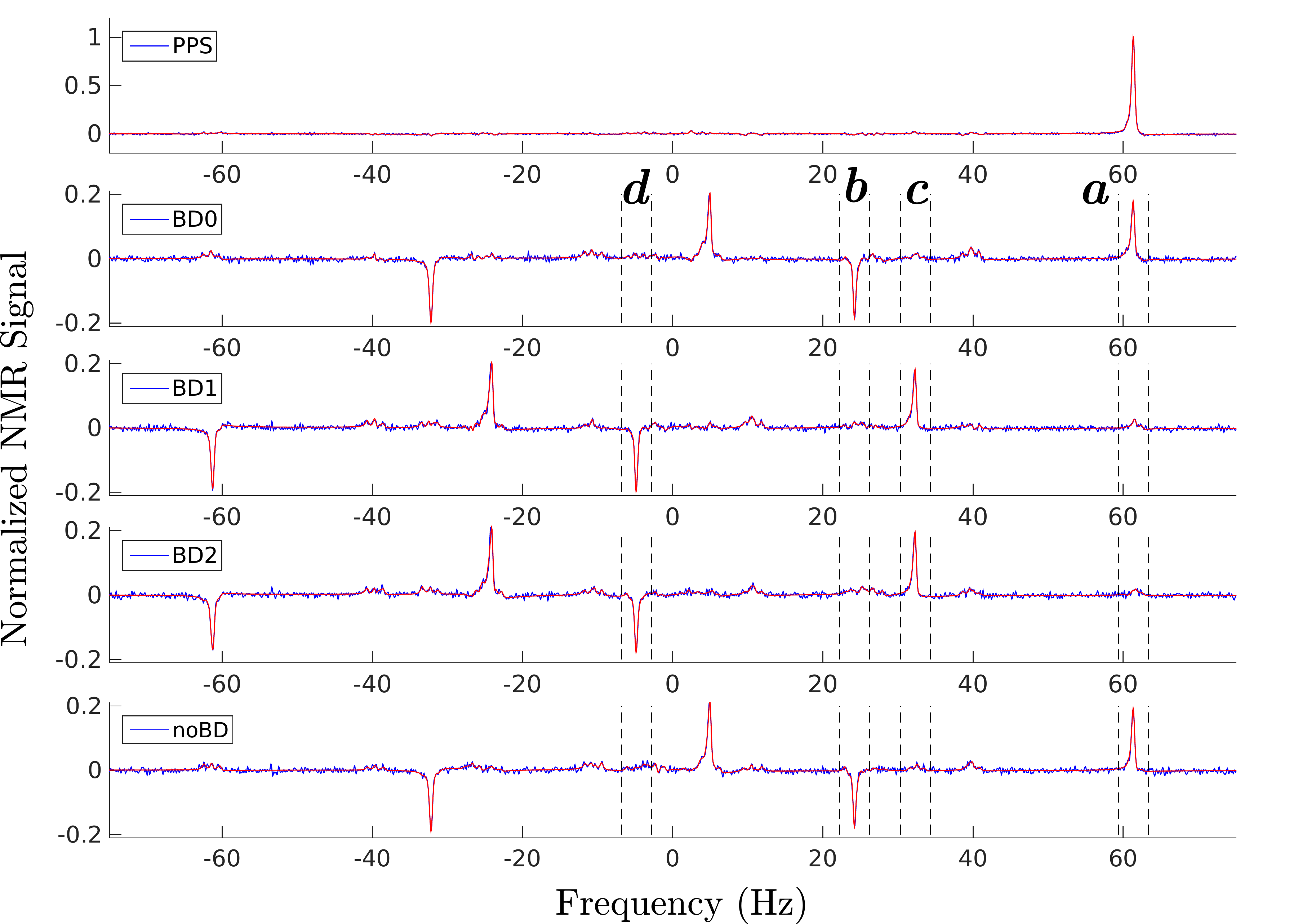}
\caption{NMR spectra of C$_7$ after the labeled-PPS, BD0, BD1, BD2, and noBD cases. The experimental data are shown in blue, and the red spectra are the fit of the experimental spectra produced by the least-square method. The labeled-PPS state shows a single peak at the expected frequency. In theory, the PPS peak splits into the four distinct peaks with equal heights for BD0, BD1, BD2 and noBD cases, and the experimental spectra show that indeed the expected four peaks appear for all cases. However, the peak height is less than a quarter of the PPS peak due to decoherence effect. As expected, for the BD0 and noBD cases, the peaks at $a$ and $b$ are more dominant than peaks at $c$ and $d$. Whereas for the BD1 and BD2 cases, the peaks at $c$ and $d$ are more dominant than peaks at $a$ and $b$.}
\label{fig:ppsbd_fit}
\end{figure*}

We measured the anyonic phases of the four different cases:
\begin{enumerate}
\item \textbf{BD0:} PPS $\rightarrow$ GD $\rightarrow$ BD0 $\rightarrow$ MM $\rightarrow$ Readout
\item \textbf{BD1:} PPS $\rightarrow$ GD $\rightarrow$ BD1 $\rightarrow$ MM $\rightarrow$ Readout
\item \textbf{BD2:} PPS $\rightarrow$ GD $\rightarrow$ BD2 $\rightarrow$ MM $\rightarrow$ Readout
\item \textbf{noBD:} PPS $\rightarrow$ GD $\rightarrow$ MM $\rightarrow$ Readout
\end{enumerate}
where noBD and BD0 ideally have $\theta_{a}=0$, and  BD1 and BD2 have $\theta_{a}=\pi$. GD and MM refer to the ground state preparation and measurement steps, respectively. Fig. \ref{fig:ppsbd_fit} shows the C$_7$ spectra of the labeled PPS and the above four cases. The experimental spectra agree qualitatively with our theoretical predictions. First, in theory, we expect to observe the same spectra for the noBD and BD0 cases and the same spectra for the two non-trivial braiding cases (BD1 and BD2) due to the path independent nature. From Fig. \ref{fig:ppsbd_fit}, it is clear that the spectra of noBD and BD0 match well, and also that BD1 and BD2 match well. Second, our experimental spectra matched well with the simulated spectra. In theory, the spectra resulting from the four cases are expected to show four peaks with equal height (two generated from $\ket{\psi_{g_{7}}}\bra{\psi_{g_{7}}}$ and the other two from $\ket{\psi_{\widetilde{g_{7}}}}\bra{\psi_{\widetilde{g_{7}}}}$), which is a quarter of the labeled-PPS peak. The spectra shown in Fig. \ref{fig:ppsbd_fit} qualitatively illustrate the expected behaviours. Third, recalling Eq. \ref{eq:phase}, we expect to observe no peaks at $c=32.24$Hz and $d=-4.93$Hz for noBD and BD, resulting in $\theta_{a}=0$, and no peaks at $a=61.25$Hz and $b=24.09$Hz for BD1 and BD2, resulting in $\theta_{a}=\pi$.  It should be noted that the other large peaks located not at the four frequencies in the spectra result from $\ket{\psi_{\widetilde{g_{7}}}}$ and are neglected in the analysis.

We estimated $|\alpha|^{2}$  and $|\beta|^{2}$ by evaluating the intensities of the peaks at the frequencies of $a$ and $b$, and frequencies $c$ and $d$, respectively. The intensities of peaks at $a$ and $b$ are averaged to estimate $|\alpha|^{2}$, and the peaks at $c$ and $d$ are averaged to estimate $|\beta|^{2}$. To evaluate the numbers, we fitted the spectra with a Lorentzian function of 64 peaks (the maximum number of observable peaks on C$_7$) using the least-square method. The experimental results of $|\alpha|^{2}$, $|\beta|^{2}$ and $\theta_{a}$ are displayed in Table \ref{table:phase} for all noBD, BD0, BD1 and BD2 cases.

The experimental results show that the anyonic phases under the two different non-trivial braiding paths $l_{1}$ and $l_{2}$ agree within the errors: (153.9$\pm$3.8)$^{\circ}$ and (151.4$\pm$3.8)$^{\circ}$. These experimental values clearly demonstrate path independence and the phase gained under the non-trivial paths compared to the cases of the trivial and no braiding paths [(17.4$\pm$6.0)$^{\circ}$ and (12.1$\pm$9.5)$^{\circ}$, respectively]. However, the experimental $\theta_{a}$ have discrepancies with the theoretical values, which are 0$^\circ$ for the trivial and no braiding paths, and 180$^\circ$ for the two non-trivial paths. For the non-trivial cases, this deviation is mostly attributed to the tiny peaks at $a$ and $b$ (Fig. \ref{fig:ppsbd_fit}), which result in $|\alpha|^{2}\approx 0.05$, because $\theta_{a}$ is highly sensitive to $|\alpha|^{2}$ as it is small and in the denominator (Eq. \ref{eq:phase}). For instance, consider a theoretical case when $|\alpha|^{2}$ is 0. In this case, regardless of a value of $|\beta|^{2}$, $\theta_{a}$ is always $\pi$. Similarly, for the trivial and no braiding cases, the deviation of $\theta_{a}$ is mostly caused by the tiny peaks at $c$ and $d$ (Fig. \ref{fig:ppsbd_fit}), resulting in $|\beta|^{2}\approx 0.02$ rather than the theoretical value of 0.
	
To investigate how the unwanted small peaks arise, we numerically simulated the NMR circuit starting from the ideal labeled PPS state using 99\% fidelity unitaries calculated from the GRAPE pulses in the presence of the decoherence effect. The assumptions that we used to simulate decoherence are shown in Appendix. The results of the simulation indicate that the errors increase the trivial loop phases to $\sim$20$^{\circ}$ whereas the non-trivial loop phases decrease to $\sim$160$^{\circ}$, blurring the difference between the two. It should be noted that most of the phase deviation comes from the decoherence effect; simulating only the gate imperfection from 99\% fidelity unitaries results in the non-trivial phases of $\sim$177$^{\circ}$. Now we discuss the different sources of error in detail.

First, the error primarily comes from the decoherence effect, and the ground state and measurement pulses contribute the most in causing the biases in the $\theta_{a}$ determination. In particular, the ground state we prepared was the ground state of the 7-qubit KM, not a ground state of our physical system. Therefore, the ground state preparation step is susceptible to decoherence, as there is no protection of the ground state by the energy gap in our NMR system.

Second, to a much lesser extent, Eq. \ref{eq:phase} no longer accurately determines the anyonic phase in the presence of gate imperfections. Therefore, to estimate the anyonic phase independent of imperfections of ground state and the measurement pulses, a different equation is required. However, it is difficult to find such an equation that is accurate and whose variables can be easily measured. Since the braiding operation is 1 ms, whereas the ground state and measurement pulses are 60 ms, the ground state and measurement pulse imperfections contribute more significantly to the $\theta_{a}$ determination. Moreover, we expect that gate imperfections are worse in experiments than in simulation, which could explain the $<$10$^\circ$ discrepancy between the simulation and experimental values after accounting for the other sources of error.

Third, we also examined the effect of $\ket{\psi_{\widetilde{g_{7}}}}\bra{\psi_{\widetilde{g_{7}}}}$ on the $\theta_{a}$ determination through numerical simulations. We simulated two scenarios with one started from $\ket{00\cdots0}\bra{00\cdots0}$ and the other from the labeled PPS. As mentioned above, the one starting with the labeled PPS results in the non-trivial $\theta_{a}$ of $\sim$160$^{\circ}$, whereas the one started from the pure state results in $\sim$150$^{\circ}$. This signifies that the contribution from $\ket{\psi_{\widetilde{g_{7}}}}\bra{\psi_{\widetilde{g_{7}}}}$ cannot be neglected completely when both gate imperfections and decoherence effects are present.

\section{Conclusions}

We have successfully demonstrated path independence of anyonic braiding statistics by braiding two anyons under two different non-trivial paths in a 7-qubit NMR quantum simulator. The anyonic phases of the two non-trivial paths $l_{1}$ and $l_{2}$ agree within the errors: $(153.9\pm3.8)^{\circ}$ and $(151.4\pm3.8)^{\circ}$ for $l_{1}$ and $l_{2}$, respectively. As references, the cases of no braiding and braiding along a trivial path are also implemented. We measured significantly smaller phases for these trivial cases compared to the non-trivial cases, confirming the extra phase acquired by the anyons in the non-trivial cases. The deviation of the anyonic phases from the theoretical value are well accounted for by the inherent errors of decoherence and imperfect gates. These contributions can be mostly attributed to the ground state preparation and measurement steps, as these steps are significantly longer than the braiding step. Other experimental schemes or setups where such a long preparation step can be prevented may be less prone to such errors. Moreover, the measurement step which is used to remarkably reduce the number of experiments in our NMR system may be eliminated in other settings.

As a step towards the realization of topological quantum computing, we do not simulate the many-body interactions in the KM Hamiltonian but alternatively use a state preparation approach to simulating the KM. This method is sufficient to simulate some particular anyonic properties such as the path independent nature shown in this paper; however, realizing fault-tolerant topological quantum computation would ultimately require engineering such Hamiltonians with many-body interactions. Fortunately, quantum simulation provides exponential speedup, outperforming classical computers as well as highly controllable systems instead of the natural intractable solid state systems. Hence, quantum simulation is a promising solution for creating and engineering the full KM Hamiltonians \cite{negrevergne2005liquid,zhang2011experimental,bloch2012quantum,franchini2014local} in the near future, and it may shed light on the goal of building a topological quantum computer in a fault-tolerant manner.

\section*{Acknowledgements}
We thank Aharon Brodutch, Jonathan Baugh, Guanru Feng and Hemant Katiyar for helpful discussions and comments, and Anthony P. Krismanich, Ahmad Ghavami, and Gary I. Dmitrienko for synthesizing the NMR sample. This work is supported by Industry Canada, NSERC and CIFAR.

\section*{Appendix A: Sample and Initialization} \label{appen:exp}

\begin{figure*}[htb]
\begin{center}
\includegraphics[width=1.4\columnwidth]{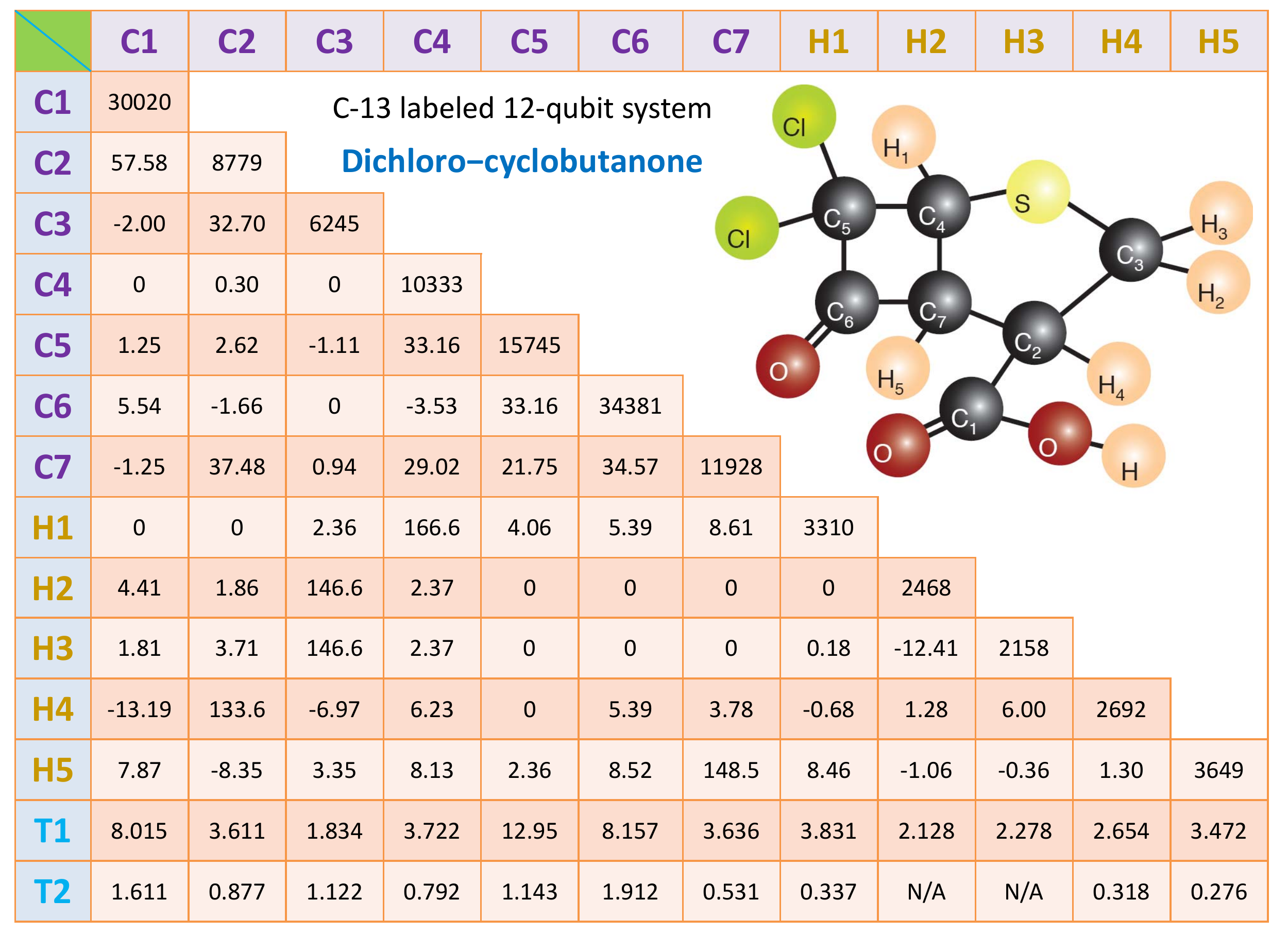}
\end{center}
\caption{Molecular structure of Dichloro-cyclobutanone, where C$_1$ to C$_7$ form a 7-qubit system. The diagonal elements are chemical shifts (Hz), and the off-diagonal elements are scalar coupling strengths (Hz). T$_1$ and T$_{2}$ are the relaxation times (Second) of the individual spins, respectively. All parameters are obtained on a Bruker DRX 700 MHz spectrometer at room temperature.}
\label{fig:mlcl}
\end{figure*}

Our NMR quantum processor is a racemic mixture of per-$^{13}$C labeled (1S,4S,5S)-7,7-dichloro-6-oxo-2-thiabicyclo[3.2.0]heptane-4-carboxylic acid and its enantiomer.  The unlabeled compound was synthesized previously by us and its structure was established unambiguously by a single crystal X-ray diffraction study \cite{johnson2008cyclobutanone}. By decoupling the $^1$H channel throughout the experiment, this sample can be regarded as a 7-qubit quantum processor which involves seven $^{13}$C spins. The $\nu_{i}$ and $J_{i,j}$ values in Eq. \ref{eq:Hnmr} of the natural Hamiltonian, as well as the relaxation time scales  T$_{1}$ and T$_{2}$, are shown in Fig. \ref{fig:mlcl}.

\begin{figure*}[htb]
\begin{center}
\includegraphics[width=1.9\columnwidth]{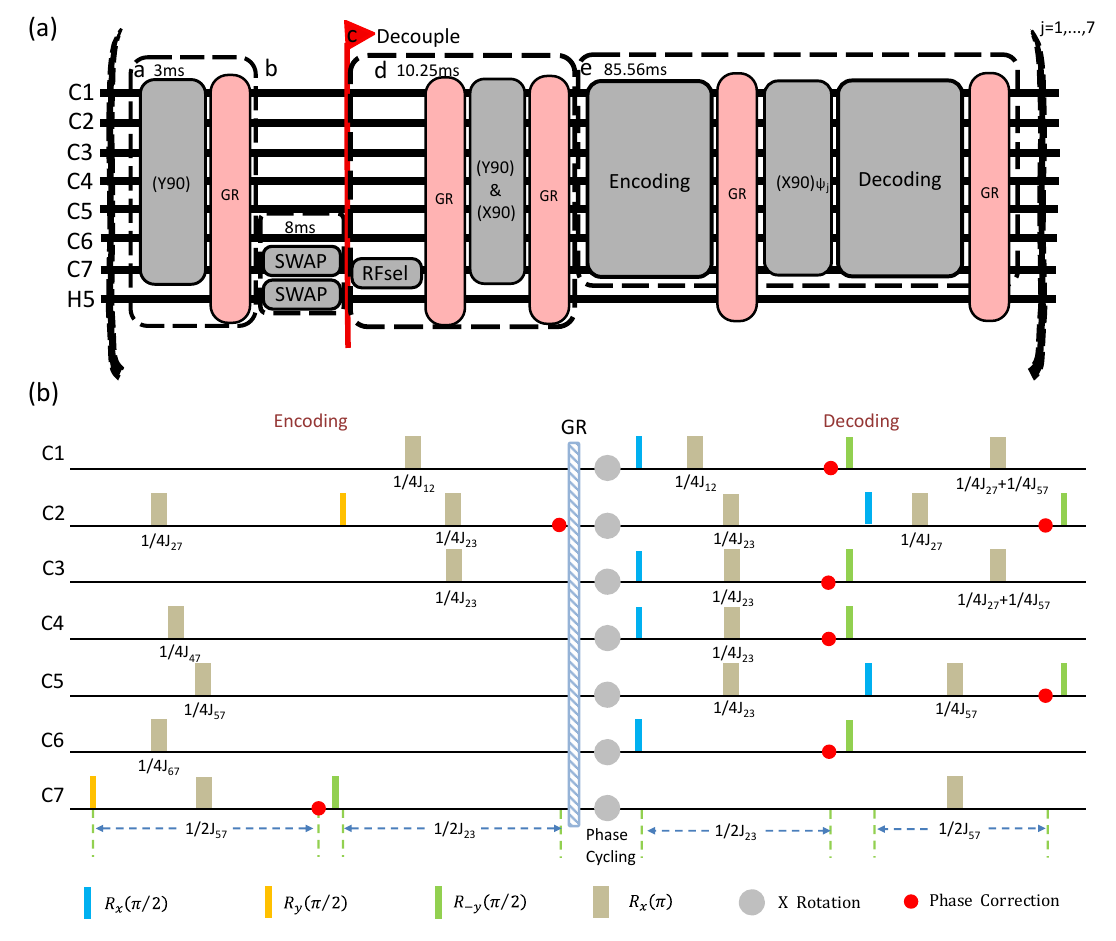}
\end{center}
\caption{(a) NMR circuit of the labeled PPS preparation. $a$: the polarization crusher step which takes the thermal state $\sum^{i=7}_{i}Z_{\text{C}_{i}}+4\sum_{i}^{i=5}Z_{\text{H}_{i}}$ to $4\sum_{i}^{i=5}Z_{\text{H}_{i}}$ by rotating all the carbon spins by $\pi/2$ around the $y$-axis (Y90) followed by a gradient field (GR); $b$: boosting the polarization of C$_{7}$ by exchanging the state of C$_{7}$ and H$_{5}$ (SWAP); $c$: decoupling the $^1$H channel for the rest of experiments; $d$: the RF-selection targeted on C$_{7}$; and $e$: labeled PPS preparation. The above steps are repeated for seven times with different phases of $\psi_{j}$ to select the appropriate coherence. For simplicity, the rest of hydrogen spins are not shown in the figure. (b) Detailed sequence of step $e$ in the above circuit.}
\label{fig:NMR circ}
\end{figure*}

We initialized the thermal equilibrium to the labeled PPS using the NMR circuit shown in Fig. \ref{fig:NMR circ}(a), where the entire circuit can be divided into five sections a-e. The input state of this 12-qubit system is the thermal equilibrium state
\bea
\rho_0 = \frac{1-\epsilon}{2^{12}}\mathbb{I}+\epsilon\left(\gamma_{\text{C}}\sum_{i=1}^7Z_{\text{C}_i}+\gamma_{\text{H}}\sum_{i=1}^5Z_{\text{H}_i}\right),
\eea
where $\gamma$ is the gyromagnetic ratio of the nuclear spins, $\mathbb{I}$ is the $2^{12}\times 2^{12}$ identity matrix, and $\epsilon\approx 10^{-5}$ represents the polarization of the system. Typically, $\gamma_{\text{C}}=1$ and $\gamma_{\text{H}}=4$ with a constant factor ignored. As the large identity matrix part does not evolve under unital operators (which is roughly the case in our experiment as the experimental time is far less than T$_1$) and it cannot be measured in NMR experiments, we can simply neglect the identity part and rewrite the input state as
\bea
\rho_0 = \sum_{i=1}^7Z_{\text{C}_i}+4\sum_{i=1}^5Z_{\text{H}_i}.
\eea
In the following calculations we only focus on this deviation density matrix assuming that the identity has no influence on the entire experiment.

\emph{a}. Rotate $^{13}$C to $\sum_{i=1}^7X_{\text{C}_i}$ by a 1 ms $\pi/2$ GRAPE pulse around $y$-axis on $^{13}$C channel, and then crush it by a 2 ms gradient pulse. The total length is 3 ms and the state at step $a$ is $\rho_a = 4\sum_{i=1}^5Z_{\text{H}_i}$.

\emph{b}. SWAP the signal of C$_7$ and H$_5$ by applying a 8 ms GRAPE pulse. This GRAPE pulse was designed via state-to-state approach and hence not a universal SWAP gate. The reason of implementing this SWAP operation is to improve the C$_7$ signal by four times in principle, which enables a much better signal-to-noise ratio (SNR) in experiment. The state at step $b$ is $\rho_b = 4Z_{\text{C}_7}+4\sum_{i=1}^4Z_{\text{H}_i}$.

\emph{c}. Turn on the Waltz-16 decoupling sequence on $^1$H channel. It averages out the signals of all $^1$H spins and their interactions with the $^{13}$C spins. In quantum information, this step is equivalent to reducing the 12-qubit system to 7 qubits which only involve $^{13}$C spins. Hence, the state at step $c$ is $\rho_c = 4Z_{\text{C}_7}$. Compared to the input thermal equilibrium state of $\rho_0$, the signal of C$_7$ has been boosted by four times.

\emph{d}. RF-selection technique is used to pick out a sub-sample which has much better RF homogeneity. As the sample in NMR has some volume in centimeters, the RF pulse applied to the sample may have inhomogeneity. Some molecules located in the centre of the RF coil experience the ideal RF amplitude, while majority of molecules experience over-rotation or less-rotation for the sake of RF amplitude inhomogeneity along the sample size. Since NMR readout is an ensemble average, the large portion with bad homogeneity contributes a lot to the final signal and causes accumulated errors when multiple pulses are implemented. RF-selection sequence \cite{Ryan2009} is such a technique to randomize this inhomogenous  portion to $x-y$ plane while keeping the homogenous portion in the thermal equilibrium state, followed by a gradient pulse in $z$-direction to destroy all $x-y$ plane signals. It is usually applied before the primary circuit, and the inhomogenous portion will stay at no-signal case during the following pulse sequence. A typical RF-selection sequence with 64 loops is
\bea
R_x(\pi/2)\left[R_{-x}(\pi)\right]^{64} \left[R_{\phi_i}(\pi)R_{-\phi_i}(\pi)\right]^{64}R_y(\pi/2),
\eea
where $\sum_i \phi_i = \pi/8$. When the molecules feel perfect RF amplitude, their states remain as thermal equilibrium after this sequence. By contrast, when the molecules feel for example 4.5\% error in RF amplitude, their states mostly evolve to $x-y$ plane and thus be killed by the following gradient field. Note that although RF-selection enables a better SNR as the RF pulses are much more precise, the cost of this technique is the absolute loss of signal as many molecules have no contributions to the signal any longer.

In our experiment, we used a GRAPE pulse instead of the long sequence to realize this RF-selection technique. This GRAPE pulse was designed on a single-qubit system via the state-to-state approach, by setting two constraints: evolve $Z$ to $x-y$ plane when the RF inhomogeneity is more than 1\%, or else do nothing to $Z$. After applying this GRAPE pulse on our 7-qubit system, we found the signal reduced to about 30\% but the RF pulses were indeed much more homogeneous by running the Rabi oscillation experiment. The two gradients and $\pi/2$ rotations in step $d$ are used to kill the minor signal of multi-coherence generated by the J-coupling evolution during the RF-selection sequence. The state at step $d$ is the same as step $c$, but with some loss that $\rho_d = 30\% \times 4Z_{\text{C}_7}$. For convenience, we simply mark this state as $Z_{\text{C}_7}$. Compared to the original thermal equilibrium state, this new state gains signal boost from H$_5$ and owns much better RF homogeneity.

\emph{e}. The main body of cat-state method \cite{Knill2000} is implemented which creates the labeled PPS $\widetilde{\rho}_{\text{PPS}}=\ket{000000}\bra{000000}Z_{\text{C}_{7}}$ from $Z_{\text{C}_7}$. It consists of three steps: encoding, phase cycling, and decoding. The detailed NMR sequence is shown in Fig. \ref{fig:NMR circ}(b). Starting from $Z_{\text{C}_7}$, the system evolves to
$Z_{\text{C}_1}Z_{\text{C}_2}...Z_{\text{C}_7}$ after the encoding step. The phase cycling step contains seven loops, and in each loop $k$ the axis of the rotation is chosen as $\text{cos}(2k\pi/7)X+\text{sin}(2k\pi/7)Y$ (the rotating angle is always $\pi/2$). The state after the phase cycling is $(\ket{00...0}\bra{11...1}+\ket{11...1}\bra{00...0})/\sqrt{2}$. The decoding step is just the inverse of the encoding part and simplified according to our molecular information. The final state after the decoding step is $\rho_e = \ket{000000}\bra{000000}Z_{\text{C}_{7}}$.

Till now the labeled PPS $\widetilde{\rho}_{\text{PPS}}=\ket{000000}\bra{000000}Z_{\text{C}_{7}}$ has been successfully prepared. Regarding the performance of this state see Fig. \ref{fig:ppsbd_fit} for its NMR spectrum.

\section*{Appendix B: Assumptions used in the Simulation of Decoherence}	
\label{section:appen1}

The list below shows the assumptions we used when numerically simulating the decoherence effects.
\begin{itemize}
\item The environment is Markovian.
\item The system and the environment are uncorrelated at t=0.
\item We only considered the effect of dephasing due to T$_{2}$ effect and neglect the effect of amplitude damping, since T$_{1}$ is much larger than the circuit time.
\item The dephasing noise is independent (or uncorrelated) between the qubits. The probability of an error happening on a given qubit does not affect the probability of an error happening on other qubits.
\item When solving the master equation, we assumed that the dissipator $D$ and the total Hamiltonian $H_{\text{tot}}$ commute for short times. Therefore, the evolution of of the state was simulated in a sequence of two steps: evolution by $e^{-iH_{\text{tot}}\Delta t}$ and subsequently, dephasing for $\Delta t$, where $\Delta t$ was chosen to match the pulse discretization. The dephasing channel implements exponential decay of off-diagonal elements according to relevant linear combinations of T$_2$ values of $^{13}$C.
\end{itemize}

\section*{Appendix C: Modified Stabilizer Measurements of the labeled PPS and Ground States}\label{appen:gd}

If the pure state $\ket{0000000}\bra{0000000}$ is prepared as an initial state, the stabilizer operators ($S_{\text{pps}}$) of such a state are $Z_{1}, \ Z_{2}, \ Z_{3}, \ Z_{4}, \ Z_{5}, \ Z_{6}, \ \text{and }\ Z_{7}$.
When these stabilizer operators evolve under the ground state preparation circuit shown in Fig. \ref{fig:svmodel}, one can reconstruct the stabilizer operators of the ground state of the 7-qubit KM.
\newcommand{\subC}[2]{$#1_{\text{C}_{#2}}$}
However, since our circuit starts from $\ket{000000}\bra{000000}Z_{\text{C}_{7}}$, the $S_{\text{pps}}$ are modified to \subC{Z}{1}\subC{Z}{7}, \subC{Z}{2}\subC{Z}{7}, \subC{Z}{3}\subC{Z}{7}, \subC{Z}{4}\subC{Z}{7}, \subC{Z}{5}\subC{Z}{7} and \subC{Z}{6}\subC{Z}{7}. The expectation values of these operators are +1 as Tr($\widetilde{\rho}_{\text{pps}}S^{i}_{\text{pps}}$) =1, where $S^{i}_{\text{pps}}$ is one of the modified stabilizer operators. These operators transform to the following operators under the implemented ground state gate which is shown in Fig. \ref{fig:NMR sample}(b):

\newcommand{\stateinC}[7]{#1$_{C_{1}}$#2$_{C_{2}}$#3$_{C_{1}}$#4$_{C_{1}}$#5$_{C_{1}}$#6$_{C_{1}}$#7$_{C_{1}}$}

\begin{align}
&1.& \text{\subC{Z}{4}\subC{Z}{7}}&\xrightarrow{U_{\text{ground}}}  \qquad\qquad Z_{\text{C}_{2}}Z_{\text{C}_{4}}Z_{\text{C}_{7}}\\ \nonumber
&2.&\text{\subC{Z}{6}\subC{Z}{7}}&\xrightarrow{U_{\text{ground}}} \qquad-Y_{\text{C}_{4}}X_{\text{C}_{5}}Y_{\text{C}_{6}}X_{\text{C}_{7}}\\ \nonumber
&3.&\text{\subC{Z}{5}\subC{Z}{7}}&\xrightarrow{U_{\text{ground}}} \qquad-Y_{\text{C}_{4}}Y_{\text{C}_{5}}X_{\text{C}_{6}}X_{\text{C}_{7}}\\ \nonumber
&4.&\text{\subC{Z}{1}\subC{Z}{7}}&\xrightarrow{U_{\text{ground}}} \qquad -Z_{\text{C}_{1}}Y_{\text{C}_{4}}X_{\text{C}_{5}}X_{\text{C}_{6}}Y_{\text{C}_{7}}\\ \nonumber
&5.&\text{\subC{Z}{3}\subC{Z}{7}}&\xrightarrow{U_{\text{ground}}} \qquad -Z_{\text{C}_{3}}Y_{\text{C}_{4}}X_{\text{C}_{5}}X_{\text{C}_{6}}Y_{\text{C}_{7}}\\ \nonumber
&6.&\text{\subC{Z}{2}\subC{Z}{7}}&\xrightarrow{U_{\text{ground}}} -X_{\text{C}_{1}}Y_{\text{C}_{2}}X_{\text{C}_{3}}Y_{\text{C}_{4}}X_{\text{C}_{5}}X_{\text{C}_{6}}Z_{\text{C}_{7}} \nonumber
\end{align}

\begin{figure}[htb]
\begin{center}
\includegraphics[width=\columnwidth]{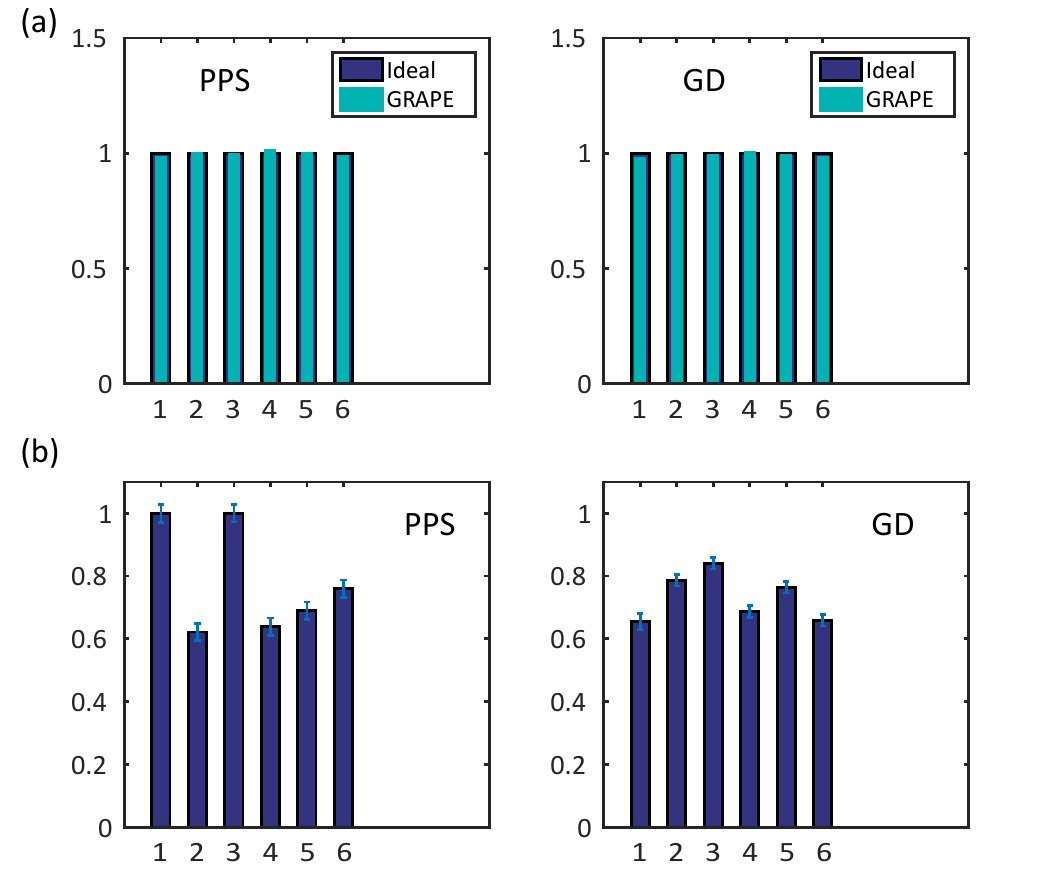}
\end{center}
	\caption{(a) Expectation values of $S_{\text{pps}}$ and $S_{\text{ground}}$. The left figure shows the expectation values of $S_{\text{pps}}$ of the theoretical and GRAPE labeled PPS state. The GRAPE labeled PPS state denotes the state numerically simulated taking the GRAPE imperfections of encoding,phase-cycling, and decoding pulses into account. Similarly, the right figure shows the expectation values of $S_{\text{ground}}$ of the theoretical and GRAPE ground states. The labeling of 1 to 6 measurements correspond to the stabilizer operators enumerated in the text of Appendix C. (b) Expectation values of $S_{\text{pps}}$ and $S_{\text{ground}}$ of the experimental labeled PPS and ground state, respectively. These values were measured by applying five different readout pulses, and the five different spectra produced from the different readout pulses which are shown in Fig. \ref{fig:gd spec} were fitted using the least-square method to estimate the coefficients of the peaks. The expectation values of the desired operators were evaluated by taking the appropriate linear combinations of the estimated coefficients \cite{Laflamme2002}. The error bars were calculated by using the method of the error propagation with the initial standard deviations from the fitting procedure.}
	\label{fig:s_theory}
\end{figure}

Therefore, the experimentally prepared ground state have +1 expectation values of the above transformed operators $S_{\text{ground}}$. We measured the expectation values of $S_{\text{pps}}$ of the labeled PPS state and the expectation values of $S_{\text{ground}}$ of the ground state. For the $S_{\text{pps}}$ measurements, a single readout pulse which rotates C$_{7}$ by $\pi/2$ around $y$-axis is sufficient to measure all six operators; whereas five different readout pulses are required (thus, five different measurements) to measure the $S_{\text{ground}}$ operators. The readout pulses are composed of the single qubit rotations that transform the product operator components of a density matrix corresponding to the $S_{\text{ground}}$ operators to the measurable product operators in C$_{7}$ spectra, which are a combination of \subC{X}{7} or \subC{Y}{7} and different \subC{Z}{i}, where $i$ indicates the $i$th $^{13}$C. For instance, the readout pulse required to measure the expectation value of the second (\#2 in the above list) $S_{\text{ground}}$ operator is $R_x^{\text{C}_{4}}(\pi/2)R_x^{\text{C}_{5}}(-\pi/2)R_x^{\text{C}_{6}}(\pi/2)$ which rotates $Y_{\text{C}_{4}}X_{\text{C}_{5}}Y_{\text{C}_{6}}X_{\text{C}_{7}}$ to $Z_{\text{C}_{4}}Z_{\text{C}_{5}}Z_{\text{C}_{6}}X_{\text{C}_{7}}$, and thus produces observable peaks at C$_{7}$ spectrum. The experimentally measured expectation values are shown in Fig. \ref{fig:s_theory}.

\begin{figure*}[htb]
\begin{center}
\subfloat[]{\includegraphics[width=\columnwidth]{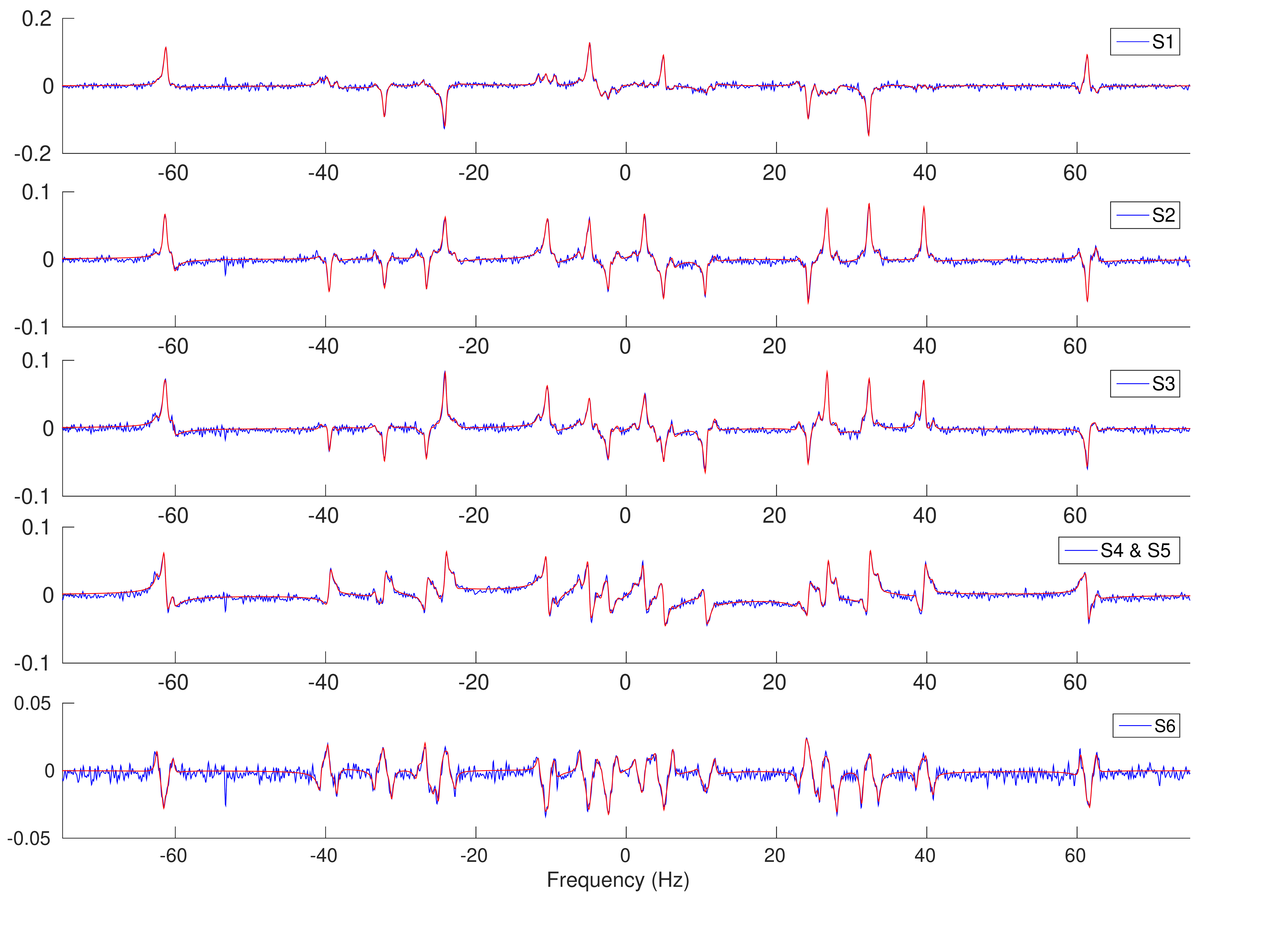}}
\subfloat[]{\includegraphics[width=\columnwidth]{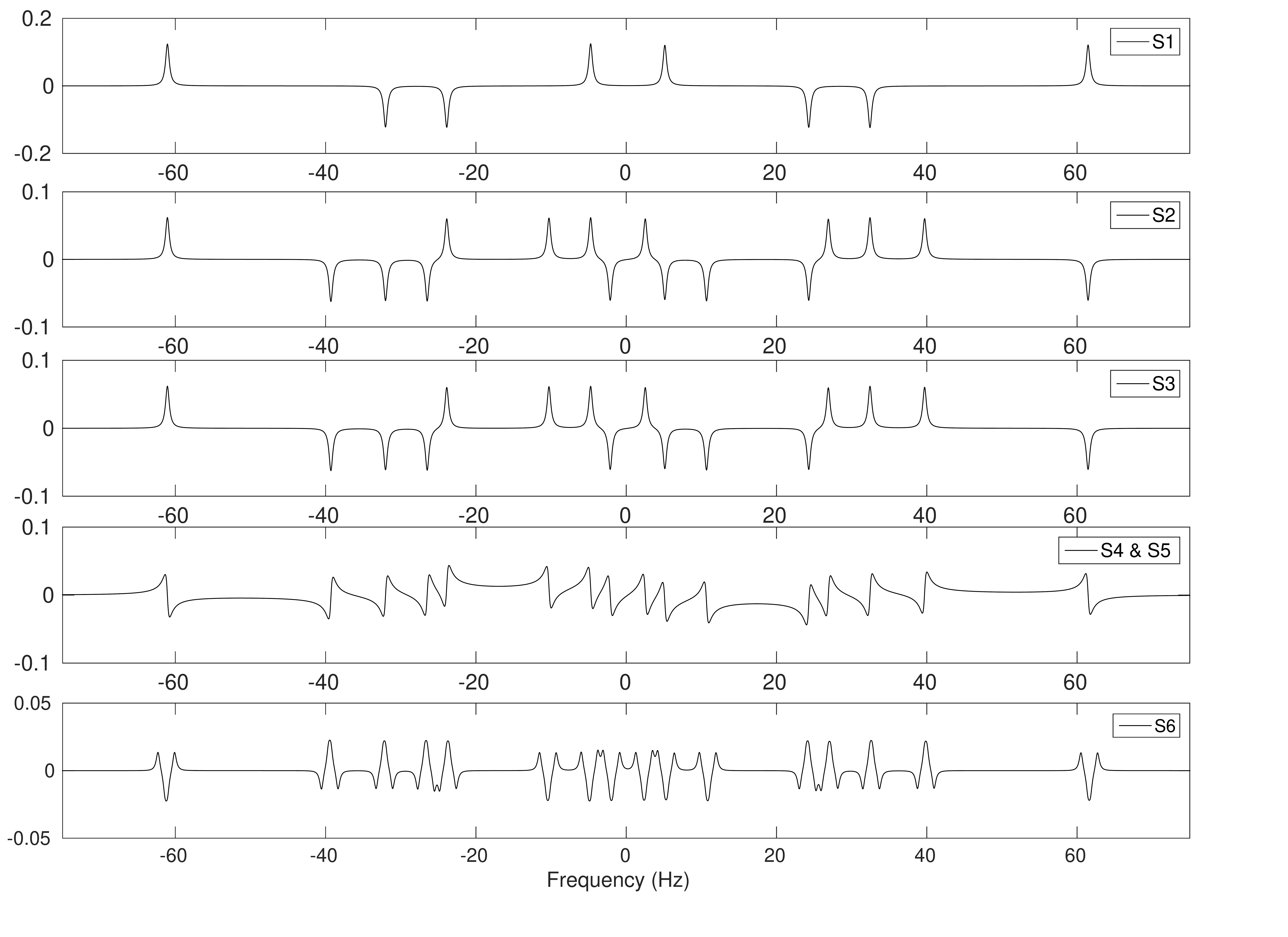}}
\end{center}
\setlength{\abovecaptionskip}{-0.2cm}
\caption{NMR Spectra used to estimate the expectation values for the stabilizer operators. (a) Experimental spectra (blue) and the fit (red) which were achieved by least-square fitting method. The legend shows which stabilizer(s) operator was(were) estimated for each spectra. (b) Theoretical spectra which were numerically simulated. Comparing (a) and (b) reflects that experiment and theory agree well qualitatively.}
\label{fig:gd spec}
\end{figure*}

Figure \ref{fig:gd spec} shows the experimental and simulated spectra of C$_{7}$ after the ground state preparation which were measured by the five different readout pulses. These spectra were used to estimate the $S_{\text{ground}}$ operators.

\end{document}